\documentclass[11pt,a4paper]{article}
\pdfoutput=1

\usepackage{jheppub}
\usepackage{amsmath,amsfonts,mathrsfs,graphicx,bbm}

\def\bee{\begin{enumerate}}
\def\eee{\end{enumerate}}
\def\bei{\begin{itemize}}
\def\eei{\end{itemize}}

\newcommand{\E}{\mathcal E}

\def\tH{\widetilde{{\cal E}}}

\def\sl(2){\alg{sl}(2)}

\def\be{\begin{equation}}
\def\ee{\end{equation}}

\newcommand{\bea}{\begin{eqnarray}}
\newcommand{\eea}{\end{eqnarray}}
\newcommand{\beml}{\begin{multline}}
\newcommand{\eeml}{\end{multline}}

\def\a {\alpha}

\def\g {\gamma}

\def\la{\label}

\def\ov{\over}

\def\S{\Sigma}

\newcommand{\alg}[1]{\mathfrak{#1}}
\newcommand{\su}{\alg{su}}

\newcommand{\AdS}{{\rm  AdS}_5\times {\rm S}^5}

\newcommand{\bem}{\left (\begin{matrix}}
\newcommand{\eem}{\end{matrix} \right )}

\def\S{{\cal S}}

\def\cO{{\mathcal O}}

\def\hstar{\,\hat{\star}\,}
\def\cstar{\,\check{\star}\,}

\def\({\left(}
\def\){\right)}

\def\eps{\epsilon}

\usepackage{hyperref}
\hypersetup{plainpages=false}

\title{Exceptional Operators in ${\cal N}=4$ super Yang-Mills}

\author[a,1]{Gleb Arutyunov,}
\note{Correspondent fellow at Steklov
Mathematical Institute, Moscow.}
\author[b,1]{Sergey Frolov}
\author[a]{and Alessandro Sfondrini}
\affiliation[a]{Institute for Theoretical
Physics and Spinoza Institute, Utrecht University, \\ Leuvenlaan \ 4, 3584 CE
Utrecht, The Netherlands}
\affiliation[b]{Hamilton Mathematics Institute and School of Mathematics, \\
~~Trinity College, Dublin 2, Ireland}
\emailAdd{G.E.Arutyunov@uu.nl, frolovs@maths.tcd.ie, A.Sfondrini@uu.nl}

\abstract{We consider one particularly interesting class of composite gauge-invariant operators in ${\cal N}=4$ super Yang-Mills theory. An exceptional feature of these operators is that in the Thermodynamic Bethe Ansatz approach the one-loop rapidities of the constituent magnons are shown to be exact  in the 't Hooft coupling constant. 
 This is used to  propose the mirror TBA description for these operators. The proposal is shown to pass
several non-trivial checks.  
}

\begin{document}

\begin{flushright}\small{ITP-UU-12-22\\SPIN-12-20
\\
TCD-MATH-12-05\\
HMI-12-02
}\end{flushright}

\maketitle

\renewcommand{\thefootnote}{\arabic{footnote}}
\setcounter{footnote}{0}

\numberwithin{equation}{section}


\renewcommand{\thefootnote}{\arabic{footnote}}
\setcounter{footnote}{0}

\section{Introduction and summary}
The aim of this work is to provide the mirror TBA description of one particularly interesting 
class of composite gauge-invariant operators in planar ${\cal N}=4$ super Yang-Mills (SYM) theory and thus to  further advance understanding of
the planar AdS/CFT \cite{Maldacena} spectral problem.

\smallskip

The operators we are interested in belong to the so-called $\su(2)$ sector of the ${\cal N}=4$ SYM and they are eigenstates of the one-loop
dilatation operator having the following explicit form \cite{BDS}
\bea
\label{Op}
{\cal O}_L= \sum_{i=1}^{L-4} (-1)^i\,{\rm tr}\Big( XX\,Z^{i}X\,Z^{L-i-3}\Big)\, .
\eea
Here $X$ and $Z$ are complex scalars of ${\cal N}=4$ SYM and $L\geq6$ is an even number. 

\smallskip

Our special interest in this class of operators 
is motivated by the following. At one loop operators from the $\su(2)$ sector can be identified with excitations of the XXX 
Heisenberg spin chain \cite{Minahan:2002ve}.
From this point of view, the operators above represent three-particle (magnon) states, and the simplest of them is an excitation of the spin chain of length 
$L=6$. Diagonalizing the Heisenberg Hamiltonian for this case, one finds the corresponding eigenvalue to be $\frac{3\lambda}{4\pi^2}$, where $\lambda$
is the 't Hooft coupling.  
Thus this state is in the spectrum of the XXX model and the same conclusion holds for all ${\cal O}_L$. However, trying to describe these states by solving the corresponding Bethe Ansatz equations one encounters a problem -- the magnons must have their rapidities $u_j$ at distinguished positions in the complex plane, namely at\footnote{Here Bethe roots are rescaled by a factor $1/2$ in 
comparison to the XXX standard normalization.} $-i,0,i$ \cite{Bethe:1931hc,BMSZ}. As a result, 
the scattering matrices entering the Bethe Ansatz are singular and the energies of such states are ill-defined\footnote{At one loop one can use
Baxter's Q-operator to describe the corresponding states in terms of dual roots which lead to the well-defined energy.}. 
This problem is, of course, well known and one natural way to cure it is to introduce a regularization by means of a twist, which we call $\phi$.  In the gauge theory twisting can be linked to the Leigh-Strassler deformation of ${\cal N}=4$ super Yang-Mills theory \cite{Leigh:1995ep} dual to strings in the Lunin-Maldacena background \cite{Lunin:2005jy} with a real deformation parameter and their nonsupersymmetric generalizations \cite{Frolov:2005dj}.
In this physical theory the limit $\phi \to 0$ can be taken without any problem. In the Bethe Ansatz approach one first computes the energy of ${\cal O}_L$ for finite $\phi$ and then takes $\phi\to 0$ finding the same result as from the direct diagonalization of the Hamiltonian.

\smallskip
 Also, having rapidities of two magnons at singular points $\pm i$ can be related to the fact that ${\cal O}_L$ is a mixture of operators where two fields $X$ are stuck together. In the terminology of \cite{Bazhanov:2010ts} two magnons form an infinitely tight bound state. We will have to say more about the nature of this bound state later.

\smallskip
Obviously, at one loop introduction of a twist is just a minor feature which distinguishes ${\cal O}_L$ from other operators.
Going to higher loops reveals more dramatic differences. To analyze the states corresponding to ${\cal O}_L$ at higher loops, we can try to employ
the all-loop asymptotic Bethe Ansatz \cite{BDS}, which is also referred to as the Bethe-Yang equations. In addition to the twist $\phi$
 the Bethe-Yang equations
depend on the coupling constant $g$ which we identify with the effective string tension related to $\lambda$ as $g=\frac{\sqrt{\lambda}}{2\pi}$.
Expanding  the Bethe-Yang equations in powers of $g$ and starting from the one-loop rapidities $0,\pm i$, one can find a formal power series solution for $u_j$ with coefficients depending on $\phi$.  As expected, nothing special happens  until one reaches the first wrapping order. However, at the first wrapping order, $g^{2L}$, one discovers that  the limit $\phi\to 0$ is singular and the corresponding energy  diverges as $\phi$ approaches zero. This behavior should be contrasted  to that  of regular states ({\it e.g.} Konishi): the latter do not even require the introduction of a twist.  On the other hand, from the point of view of the gauge theory we should do not expect any problem with taking $\phi\to 0$ for operators of the type ${\cal O}_L$. 

\smallskip

Certainly, the Bethe Ansatz is only asymptotic, that is it provides a correct description of the spectrum only up to the first wrapping order; the perturbative behavior of ${\cal O}_L$  serves as a clear confirmation of this fact.  Hence, as for regular operators, we should expect that the TBA must give an adequate solution. 

\smallskip

We recall that the TBA approach, originally developed for relativistic theories \cite{Zamolodchikov90}, enables a computation of the ground state energy
of a two-dimensional integrable model in a finite volume by evaluating the partition function of the accompanying mirror model \cite{AF07}.
In recent years the mirror TBA -- a tool to determine energies of string states on $\AdS$ and correspondingly scaling dimensions of gauge theory operators --
has been largely advanced \cite{AF09a}-\cite{Balog:2011cx}  and generalized to include excited states \cite{GKKV09}-\cite{Arutyunov:2011mk}.
Results derived from the corresponding TBA equations  \cite{GKV09b}-\cite{BH10a} show an  agreement with various string 
  \cite {Gromov:2011de}-\cite{Beccaria:2011uz} and gauge theory \cite{Sieg}-\cite{Eden:2012fe} computations, and also with L\"uscher's perturbative treatment \cite{BJ08}-\cite{Janik:2010kd}.   
 \smallskip
 
 Apparently, constructing the TBA equations for the states corresponding to ${\cal O}_L$ we might follow the same procedure as for regular states.
 This amounts to first building up an asymptotic solution with a finite twist\footnote{Introduction of a twist in the mirror TBA has been considered in the recent work  \cite{Arutyunov:2010gu}-\cite{deLeeuw:2012hp}.}, analyzing its analytic properties and then using them to engineer the TBA equations \cite{Arutyunov:2011uz}.  However, the TBA equations constructed in such a way rely on the asymptotic solution which is valid only for 
 $\lambda \lesssim \phi$ which makes obscure  how to take the limit $\phi\to 0$ with $\lambda$ fixed. 
 More precisely, for fixed $\phi$ there always exists a critical value $\lambda_{\rm cr}\equiv \lambda_{\rm cr}(\phi)$ such that the Bethe-Yang equations have a well-defined solution for $\lambda\leq \lambda_{\rm cr}(\phi)$
 and no solution for $\lambda > \lambda_{\rm cr}(\phi)$. Nevertheless, in perturbative treatment of the TBA
 this problem of order of limits can be overcome  by considering first the expansion in powers of $\lambda$ and then taking the limit $\phi\to 0$ in each term of the expansion. In this work we consider in detail the corresponding twisted TBA equations for ${\cal O}_L$ with $L=6$.  In fact, introduction of the twist results in the analytic behavior of rapidities and Y-functions very similar to that considered in \cite{Arutyunov:2011mk}, in particular, 
  the complex rapidities $u_2$ and $u_3$ of the second and third particle respectively,
lie outside the analyticity strip, which is in between two lines running parallel to the real axis
at $-\frac{i}{g}$ and $\frac{i}{g}$. Not surprisingly, the TBA equations for the state corresponding to ${\cal O}_6$ essentially coincide with that 
of   \cite{Arutyunov:2011mk}.  By expanding these TBA equations up to $\lambda^6$, we then show that  the TBA correction to the Bethe-Yang equations cancels precisely the divergent part of the asymptotic 
 energy rendering therefore the limit $\phi\to 0$ well-defined.  For the energy $E^{(6)}$ at six loops (the first wrapping order) we then find
 \bea
 \nonumber
E^{(6)}=\left( -\frac{84753}{1024} + \frac{243 }{128}\zeta(3) + \frac{189}{64}\zeta(5) - \frac{567}{128}\zeta(9)+{\cal O}(\phi)\right)g^{12}\, .
\eea 
 We also provide a mechanism for a similar cancellation at higher orders of $\lambda$.   This in principle solves the problem of describing singular states in perturbative theory. It is quite remarkable that in spite of the fact that the TBA corrections make the energy of a state finite in the limit $\phi\to 0$, the perturbative rapidities remain divergent in this limit. 
 \smallskip
 
 A veritable question is however how to describe singular states for finite $\lambda$ and what are the corresponding  TBA equations. To answer 
 this question, we again consider a state  which contains only our three distinguished magnons. For large $L$ such a state can be viewed as a scattering state of a fundamental particle and a two-particle bound state with momenta $\pm\pi$.  We put forward a conjecture that the one-loop rapidities are in fact {\it exact for any value of $\lambda$}, and we use this conjecture to propose TBA equations for these states. In what follows we refer to these rapidities as exceptional. As a very non-trivial
 consistency check, we show that our conjectured TBA equations lead to the constraints\footnote{Here $Y_{1_*}$ denotes analytic continuation of the main Y-function $Y_1$ to the string region.} 
 $Y_{1_*}(u_j)=-1$ which for regular states would have to be imposed as momentum quantization conditions.
 We  compute the energy of the shortest operator of this type (of length $L=6$)   up to $\lambda^6$ and show that 
it perfectly agrees with the result obtained from the twisted TBA equations.  We believe that the equality of energies computed from the TBA based on twisted and exceptional rapidities must hold to all orders in perturbation theory. 

\smallskip

Amazingly, in the approach based on the exceptional rapidities,
the TBA corrections begin to contribute to the energy already at $\lambda^3$, and for a generic singular state at $\lambda^{L/2}$, {\it i.e.} {\it at half-wrapping}. This behavior is consistent with the analysis of \cite{AF07} where a two-particle bound state with the total momentum $p$ larger than the critical value
$p_{\rm cr}$  has been studied. Indeed, the leading exponential correction to the energy of the bound state found from the Bethe-Yang equations is  $e^{-qJ}$ and the leading TBA correction is expected to be of the same order.
Here $q$ is used to parametrize the complex particle
momenta $p_2=\frac{p}{2}+iq$ and $p_3=\frac{p}{2}-iq$ with ${\rm Re}\, q>0$. 
At weak coupling $p_{\rm cr}\approx\pi -2 g$, {\it i.e.} the momentum $p=\pi$ of the two-particle bound state we are interested in here exceeds the critical value.  According to 
\cite{AF07}, in the limit $g\to 0$ one has $q=-\log\frac{g}{2}+\ldots$, {\it i.e.} the leading TBA correction must be of the order $e^{-qJ}\sim g^{J}$ where
$J\approx L$ is large.

\smallskip

The family of three-particle states corresponding to ${\cal O}_L$ is probably the only example of  states which rapidities are known as exact functions of $g$.
For this reason we call the operators  ${\cal O}_L$ {\it exceptional}. In a sense these states are similar to the vacuum state for which one does not have the exact Bethe equations. 
Of course, the TBA equations for ${\cal O}_L$ are non-trivial and they are ultimately responsible for the non-trivial dependence of energy on the coupling constant. It would be very interesting to see whether ${\cal O}_L$ exhibit exceptional features also from purely field-theoretic point of view. 

\smallskip

\begin{figure}[t]
\begin{center}
\includegraphics[width=0.8\textwidth]{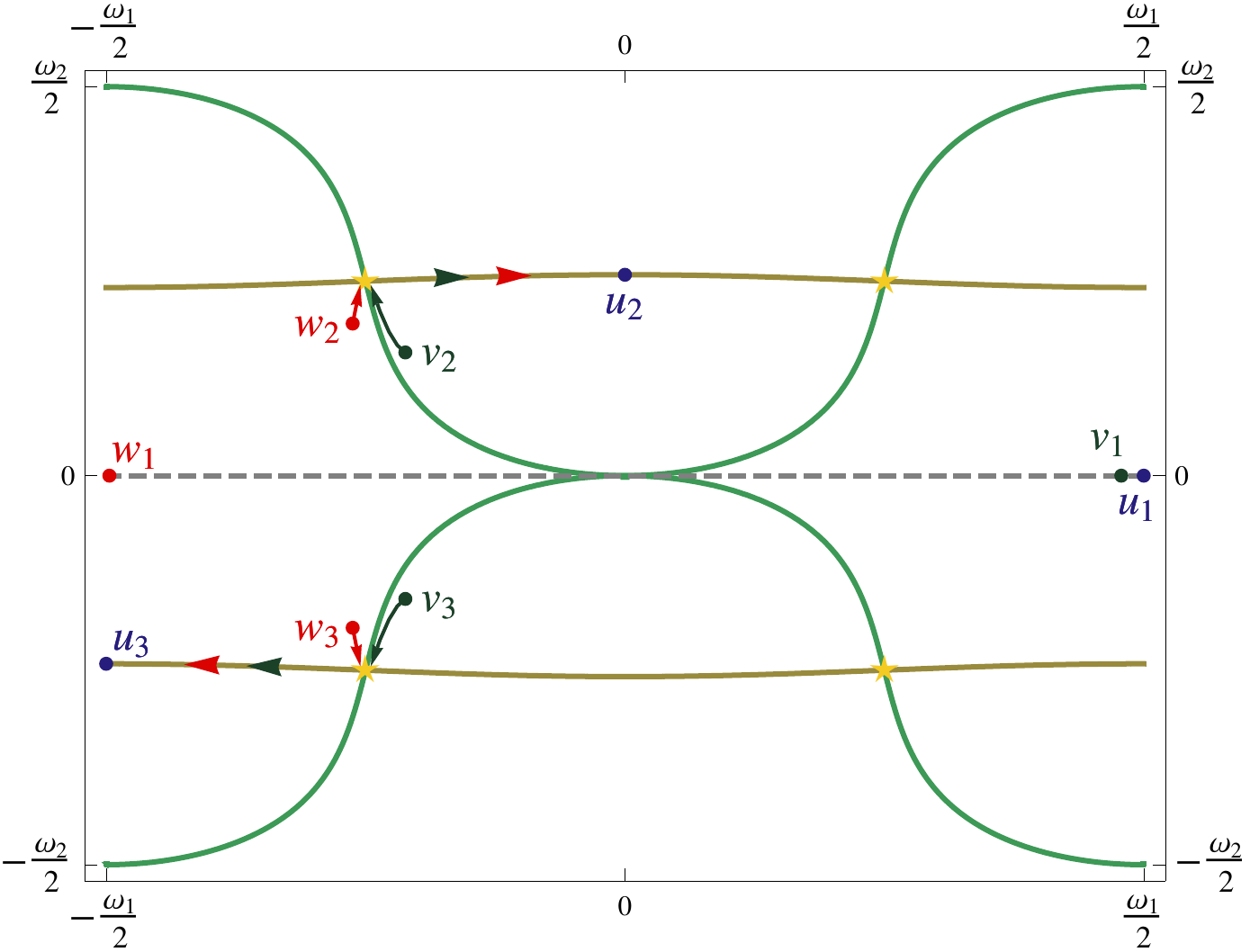}
\caption{The picture of the $z$-torus at $g=0.1$.  Brown and green curves are the boundaries of the string and (anti-)mirror regions. They intersect at four points which correspond to the branch points on the string $u$-plane.  Exceptional rapidities are $u_{1}=0$, $u_{2,3}=\mp i/g$. Twisted rapidities are indicated by $v_i$ and $w_i$.
Rapidities $v_{2,3}$ are located just a bit outside the analyticity strip as happens for {\it e.g.} $L=6$, while $w_{2,3}$ are inside as for {\it e.g.} $L=8$. Arrows indicate the conjectured motion of the twisted rapidities as $\phi\to 0$.   }
\label{trus_rapidities}
\end{center}
\end{figure}

In fact one can consider more general operators which include the three exceptional magnons as a building block \cite{BMSZ}.
In contrast to the exact rapidities of exceptional magnons, extra rapidities of such an operator are not rigid and have non-trivial $\lambda$-dependence. 
The results of this paper allow one to readily construct the corresponding TBA equations. In a sense all such states can be viewed as a new sector 
of ${\cal N}=4$ SYM with exceptional operators playing the role of non-BPS vacuum states.  

\smallskip

Having established two TBA approaches to exceptional operators -- the twisted one and the one based on the exceptional rapidities (both producing the same perturbative energies) -- one can naturally wonder what is the relation between them.   Apparently, they look rather different, in particular, in the twisted approach the perturbative rapidities are divergent in the limit $\phi\to 0$. To clarify this issue, one can fix 
a value of  $\lambda$ and look for the evolution of the rapidities $u_j(\phi)$ when the twist decreases from some finite value to zero. 
Inverting the function $\lambda_{\rm cr}(\phi)$, one finds a critical value of the twist  $\phi_{\rm cr}=\phi_{\rm cr}(\lambda)$. For $\phi>\phi_{\rm cr}(\lambda)$
the Bethe-Yang equations have a solution corresponding to exceptional operators, while as far as $\phi<\phi_{\rm cr}(\lambda)$ the solution 
ceases to exist. A characteristic property of $\phi_{\rm cr}(\lambda)$ is that it vanishes in the limit $\lambda\to 0$. Importantly, one finds that when $\phi$ approaches $\phi_{\rm cr}$ from above the complex rapidities $u_{2,3}$ move towards the branch points of  the string $u$-plane at $-2\mp i/g$,
where  the function $Y_2$ develops a double pole. On the $z$-torus the branch points correspond to the points of intersection of the boundaries of the string and (anti-)mirror region, 
see figure \ref{trus_rapidities}.   Decreasing the twist below 
$\phi_{\rm cr}$, the only way to smoothly continue the evolution of  $u_{2}$ and $u_3$ compatible with reality of the energy 
is to assume that they move along the cuts of the string $u$-plane or 
on the $z$-torus along the boundaries of the string region in opposite directions, reaching the positions of the exceptional rapidities 
at $\phi=0$. On the $z$-torus 
all the way towards the branch points the rapidities $u_2$ and $u_3$ remain complex conjugate but they loose this property upon passing them. 
On the $u$-plane this corresponds to the fact that $u_2$ and $u_3$ move along the lower edges of the cuts which reflects our choice of the string $u$-plane.
In fact, such a behavior of $u_{2,3}$ is the same as the one found in \cite{AF07} for a two-particle BPS bound state at infinite $J$. 
Concerning the divergency of rapidities in the twisted theory, it is (almost) certain that this  is just an artifact of the perturbative expansion. For finite $\lambda$ the rapidities may have an essential singularity at $\phi=\phi_{\rm cr}$ such that the limit $\phi\to 0$ would produce the exceptional rapidities we conjecture. For example a term 
$\phi\, e^{-\lambda/(\phi-\phi_{\rm cr}(\lambda))^2}$ leads to poles in $\phi$ in perturbative theory while for finite $\lambda$ it gives a zero contribution in both limits $\phi\to \phi_{\rm cr}$ and $\phi\to 0$. It would be important to further justify the above-described scenario, in particular to construct the TBA equations for $\phi<\phi_{\rm cr}$
and show their consistency with our assumptions of positioning the rapidities on the boundaries of the string region.  It is worth stressing that 
for  these rapidities the usual asymptotic description does not exist because some S-matrices are singular.  Nevertheless, the existence of the TBA  for  exceptional rapidities indicates that the corresponding construction must exist also for this case.

\smallskip

The paper is organized as follows. In the next section we discuss the emergence of singular states in the asymptotic Bethe Ansatz
and introduce a twist. For the three-magnon case and $L=6$ we also provide a perturbative solution of the Bethe-Yang equations up to the order $\lambda^6$ accompanied by a small $\phi$-expansion which reveals a singular nature of the state under consideration. In section 3
we discuss the twisted TBA equations for singular states and also compute the first L\"uscher correction to the energy for the state corresponding to ${\cal O}_6$. We then show that the energy admits a smooth limit $\phi\to 0$. To shed further light of finiteness of energy in the twisted TBA approach,
we explicitly demonstrate a cancellation of the leading singularities in the expression for the energy at order $\lambda^7$.   Section 4 is devoted to the TBA approach based on exceptional rapidities. After formulating our conjecture on the exact form of $u_j$, we analyze the analytic properties of the asymptotic Y-functions which appear to be remarkably simple. Relying on the analytic structure of the asymptotic solution, we  construct the  
corresponding TBA  equations and show that they imply the fulfillment of the exact Bethe equations.  We then compute the energy of ${\cal O}_6$ and show that in spite of the fact that in the approach based on the exceptional rapidities the TBA 
starts to contribute to the energy already at half-wrapping, the energy perfectly agrees with that found from the twisted TBA up to and including 
the first wrapping order.
In the conclusions we discuss some interesting problems for future research.
Some technical details are relegated to four appendices, and explicit expressions for twisted rapidities and Y-functions can be found in the Mathematica file attached to the arXiv submission of the paper.

\section{Bethe-Yang equations and singular rapidities}
In a perturbative expansion in $g=\frac{\sqrt{\lambda}}{2\pi}$
 wrapping effects contribute to the scaling dimension starting from order $g^{2L}$ where $L$ is the length of the operator under consideration. Consequently, the Bethe-Yang equations provide the description of the perturbative spectrum up to the first wrapping order, and its predictions are usually expected to be qualitatively true even for finite but small $g$. It is therefore natural to start our analysis of exceptional operators with the corresponding Bethe-Yang equations.

In what follows we will interchangeably use the gauge and string theory language, speaking equivalently of scaling dimension (of a gauge invariant operator) and energy (of the correspondent string excitation), etc.

\subsection{Singular rapidities in the one-loop Bethe Ansatz}
The one-loop spectrum of $\mathcal{N}=4$ SYM in the $\su(2)$ sector is described by the XXX spin chain 
\cite{Minahan:2002ve}. 
Scaling dimensions can be found by solving the Bethe ansatz equations for rapidities of $M$ magnons 
\bea
1=e^{ip_k L}\prod_{j\neq k}^MS_{\rm \scriptscriptstyle xxx}(u_k,u_j)
\quad\Rightarrow\quad
1=\left(\frac{u_k+i}{u_k-i}\right)^L\prod_{j\neq k}^M\frac{u_k-u_j-2i}{u_k-u_j+2i}\,,\quad k=1,\dots,M\, .~~~~
\eea
Invariance under cyclic permutations\footnote{In string theory this is equivalent to imposing the level-matching condition.}  implies
\bea
\la{1loopPis0}
e^{iP}=1\quad\Leftrightarrow\quad\prod_{k=1}^M\frac{i+u_k}{i-u_k}=1\,,
\quad{\rm with }\quad P=\sum_{k=1}^Mp(u_k),\quad p(u)=-i\,\log\frac{i+u}{i-u}\, .
\eea
The one-loop scaling dimensions, or energies, are then given by
 \bea
E=L+g^2\sum_{k=1}^M\frac{2}{1+u_k^2}\,.
\eea
Solutions of the Bethe-Yang equations exist also for complex values of the rapidities. It has been observed \cite{Bethe:1931hc,BMSZ} that among those there exist solutions with odd $M$ where three rapidities are placed at 
\bea
u_1=0\,,\quad \quad u_2=-i\,,\quad \quad u_3=i\,,
\eea
 and the remaining $M-3$ rapidities come in pairs. The first three rapidities are rather exceptional: the corresponding momenta read
\bea
p_1=\pi,\quad\quad p_2=-\frac{\pi}{2}+i\,\infty,\quad\quad p_3=-\frac{\pi}{2}-i\,\infty,
\eea
 and similarly the individual energy of each of the last two magnons is ill-defined, signaling the necessity to introduce a regularization. This can equivalently be done by introducing a regularization parameter $\varepsilon$ in the solutions $u_1=f_1(\varepsilon)$ and $u_{2,3}=\mp i +f_{2,3}(\varepsilon)$ as in \cite{BMSZ,BDS} or by introducing a twist $\phi$ in the Bethe-Yang equations as {\it e.g.} in \cite{Bazhanov:2010ts}:
\bea
\la{oneloopBYtwist}
1=e^{-i\phi}\left(\frac{u_k+i}{u_k-i}\right)^L\prod_{j\neq k}^M\frac{u_k-u_j-2i}{u_k-u_j+2i}\,,\quad k=1,\dots,M.
\eea
Then the cyclicity condition (\ref{1loopPis0}) becomes 
$P=M\phi/L$ mod $2\pi$.

Focusing on the case $M=3$, where only the three exceptional rapidities are present, one finds that when $L$ is even (and of course $L\geq6$) solutions can be found so that in the limit $\phi\to0$ rapidities tend to $u_1=0$ and $u_{2,3}=\mp i$.
 This can be done by requiring that the divergence of momenta for small $\phi$ is compensated by a singularity in the S-matrix $S_{\rm \scriptscriptstyle xxx}(u_2,u_3)$. Schematically one then has
\bea
\la{onelooptwistedsol}
u_1\sim\phi\,,\quad\quad u_2\sim -i-\phi-i\,\phi^L\,,\quad\quad u_3\sim +i-\phi+i\,\phi^L\,,
\eea
where the value as well as the sign of the coefficient of the imaginary correction depends on $L$. Then, for all even $L$, the scaling dimension of the operator or equivalently the energy of the dual string state is also regular and reads
\bea
\lim_{\phi\to0}E(\phi)=L+3g^2\,.
\eea
Furthermore, the corresponding one-loop eigenvectors of the dilatation operator can be found by taking the limit of the Bethe wave-function of the twisted solution, yielding the $\mathcal{N}=4$ SYM operators \eqref{Op}.
Therefore, at one loop, we conclude that there exists a family of eigenstates of the dilatation operator that can be constructed out of a building block of three exceptional magnons. These can be thought of as one magnon of maximal momentum $p_1=\pi$ and one ``infinitely tight'' two-magnon bound state having maximal momentum $p_2+p_3=-\pi$.  It is interesting to see whether and how this picture changes beyond one-loop.

\subsection{All-loop Bethe-Yang equations and their breakdown}
The all-loop Bethe-Yang equations in the $\su(2)$ sector \cite{BDS,AFS,BES}  including the twist\footnote{As discussed in more detail in appendix \ref{app:gammadeform}, the twisted Bethe-Yang equations (together with the twisted level-matching condition) describe a $\gamma$-deformation of $\mathcal{N}=4$ SYM.} read
\bea
\la{allloopBYtwist}
1= e^{-i\phi}e^{ip_kL}\prod_{j\neq k}^M \frac{u_k-u_j-2i}{u_k-u_j+2i}\sigma^{-2}(u_k,u_j)\,,
\eea
where $L=J+M$ and $\sigma(u_k,u_j)$ is the dressing factor. Here and in what follows we adopt the notation usual to field theory in which rapidities approach constant values for small $g$. Therefore,
\bea
x_k^\pm=x_s\big(u_k/g\pm i/g\big)\,,\quad\quad\,x_s(u)=\frac{u}{2}\left(1+\sqrt{1-4/u^2}\right)\,,
\eea
and the relation between rapidity and momentum of a magnon is $e^{ip_k}=x_k^+/x_k^-$. Again, the equations are supplemented by the level-matching condition 
$e^{i\,P}=e^{iM\phi/L}$.

As before, we focus on three-excitation solutions that for small $g$ tend to the one-loop configuration of the previous section. From field theory, one expects the scaling dimension of any operator to admit a well-behaved small coupling expansion. Therefore, one would hope to resolve any singularity in the Bethe ansatz description by the same means used in the previous section.

\begin{figure}[t]
\begin{center}
\includegraphics[width=0.45\textwidth]{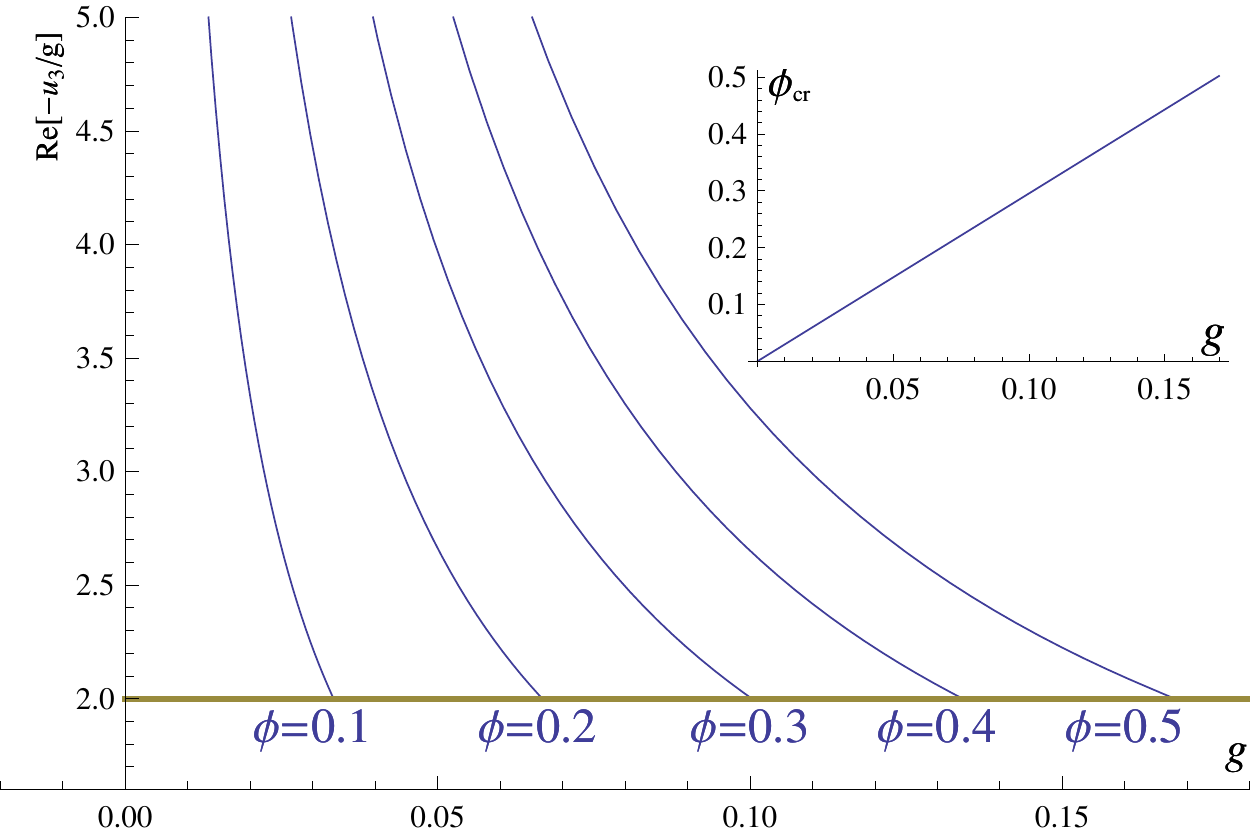}\qquad
\includegraphics[width=0.45\textwidth]{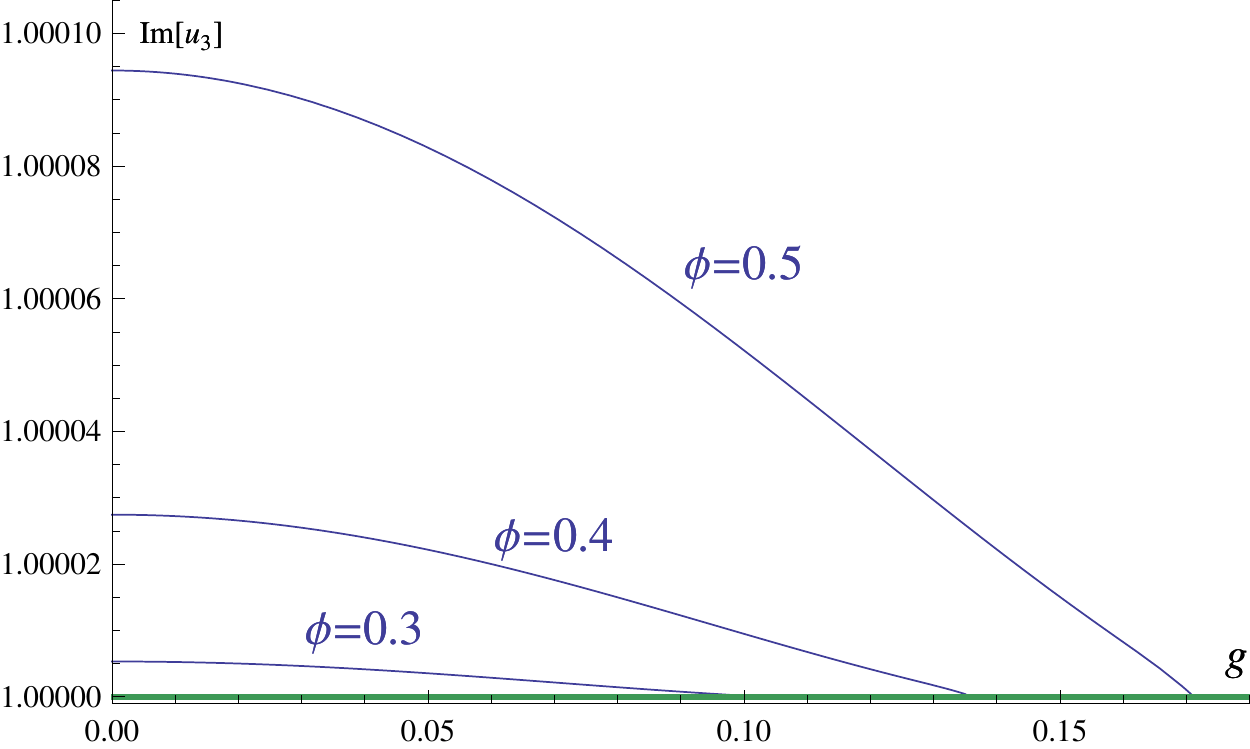}
\caption{Plots of the real and imaginary parts of $u_3$ as functions of $g$ for various values of $\phi$. For any fixed $\phi$ the rapidity $u_3$
moves to the branch point $-2g+i$ in the field theory normalization and reaches it at $g_{\rm cr}(\phi)$.  
Inset represents the inverse function $\phi_{\rm cr}(g)$ which apparently is a linear function of $g$ with slope $\approx 3$. }
\label{rapidities}
\end{center}
\end{figure}

Let us consider, for simplicity, the case of the shortest operator of length  $L=6$. Then, for any non-vanishing value of $\phi$, we can numerically solve (\ref{allloopBYtwist}). Some of these solutions are plotted in figure \ref{rapidities}. These describe one particle with real rapidity and a pair of  particles with complex conjugate rapidities for small $g$. However, as noticed in similar cases \cite{Arutyunov:2011mk}, it appears that the solution predicted by the Bethe-Yang equations breaks down at some critical value of the coupling $g_{cr}(\phi)$, which depends on the twist, see figure \ref{rapidities}.
There the rapidities are no longer complex-conjugate to each other, and as a result the energy becomes complex.

We expect the breakdown to be an artifact of the asymptotic nature of the Bethe-Yang equations. What is striking, and peculiar of these states, is that the value of $g_{cr}(\phi)$ where the breakdown happens goes to zero
 with $\phi$, and therefore for finite $g$ the twist cannot be removed no matter how small $g$ is. This scenario also holds  for larger values of $L$.

This raises the question of whether the asymptotic description can be employed at least perturbatively in $g$.  Expanding (\ref{allloopBYtwist}) perturbatively, up to the order $g^{2L-2}$ one can find a solution of the form\footnote{The solution for rapidities for $L=6$ can be found in the Mathematica file attached to the arXiv submission of this paper. }
\bea
u_i=\sum_{n=0}^{L-1}f_{i,n}(\phi,L)\,g^{2n}+{\cal O}(g^{2L})\,,
\eea
where at $\phi=0$ the coefficients $f_{i,n}(0,L)$ are regular. The energy up to ${\cal O}(g^{2L})$ is then found from the asymptotic formula 
\bea
\la{allloopdispersion}
E^{\rm asym}=J+\sum_{k=1}^M\sqrt{1+4g^2\,\sin^2(p_k/2)}\, ,
\eea
which involves the all-loop dispersion relation only.
On general grounds we expect the asymptotic formula to receive corrections at order $g^{2L}$ due to wrapping effects, and therefore to differ from the ``true'' result (which in principle might be computed by field theory perturbative techniques). For these particular states, however, the asymptotic energies appear to be \emph{divergent} in the limit $\phi\to0$ at the wrapping order. For instance, in the $L=6$ case  we find
\bea
\la{divergenEn}
\hskip -0.2cm
&&\hskip -0.2cm E^{\rm asym}=6 + 3 g^2 - \frac{9}{4} g^4 + \frac{63}{16} g^6 - \frac{621}{64} g^8 - 
 \frac{9}{256}(8\zeta(3)-783 ) g^{10}+\\
&&\hskip -0.2cm +\Big(-\frac{2187}{1024 \phi^6} - \frac{3645}{
 8192 \phi^4} + \frac{189783}{1310720 \phi^2}+ \frac{81}{128} \zeta(5) + \frac{27}{32} \zeta(3)-\frac{1223982387}{14680064}\Big)g^{12}+\cO(g^{14},\phi)
 \nonumber
\eea
Starting from the wrapping order $g^{2L}$, the rapidities also become divergent in the limit $\phi\to 0$.
This result is remarkable. Indeed, doing perturbative computations in $\gamma$-deformed $\mathcal{N}=4$ SYM one would find that for small $\phi$ the numerical discrepancy between the asymptotic prediction and the true result is enormous. 
Obviously this is related to the fact that  wrapping corrections have been neglected so far. Since the asymptotic energy diverges as $\phi$
approaches zero, contribution of wrapping diagrams becomes crucial for diagonalization of the mixing matrix.   
This means that for exceptional states (or for states containing the three exceptional rapidities) a separation of the exact energy 
into asymptotic and wrapping parts is ill-defined in the limit of vanishing twist.     

In order to properly account for wrapping effects, we will use the mirror TBA. 
 A convenient approach to excited states TBA is to make use of the contour deformation trick and  of the knowledge of analytic properties of asymptotic Y-functions. For this purpose it is convenient to formulate TBA equations in the twisted theory for $g\lesssim\phi$ where the asymptotic description can be trusted.

\section{Twisted TBA}
We want to find the mirror TBA description of the exceptional three-magnon configurations discussed in the previous section, which we expect to exist for any even $L\geq 6$. Our strategy will be to introduce a twist $\phi$ and first formulate the TBA equations for the twisted theory, which corresponds to a $\gamma$-deformation of $\mathcal{N}=4$ SYM.

Fixing a length $L$, for any nonzero $\phi$ and for $g$ small enough we can find the asymptotic solution of the twisted Bethe-Yang equations (\ref{allloopBYtwist}). These in turn allow one to write down the asymptotic Y-functions in the twisted theory. The details of this construction are given in appendices \ref{app:Yfunctions} and \ref{app:TBA}. Knowing the analytic properties of the asymptotic Y-functions,  we can write down the TBA equations, which can then be solved numerically or perturbatively in $g$.

\subsection{Analytic structure of  Y-functions } 
We are considering here a family of configurations (labeled by even $L$) with one real rapidity $u_1$ and two complex-conjugate  $u_{2,3}$, depending on $g$ and $\phi$. Since eventually we are interested in the limit $\phi\to0$, we restrict ourselves to considering a small region of parameter space,
\bea
g\lesssim\phi\ll1\,,
\eea
where the first inequality follows from the necessity of having a real energy solution of the Bethe-Yang equations.

Different states in the family have slightly different analytic structure for auxiliary Y-functions, that in turn yield different driving terms in the TBA equations by contour deformation trick. The procedure to formulate these equations in the case of complex rapidities has been detailed in \cite{Arutyunov:2011mk}, and can be applied straightforwardly to our case with minor $L$-dependent modifications.

Therefore, rather than attempting to give a unified description of each state in the family, we focus on the shortest one, with $L=6$. In order not to clutter our treatment with technicalities, we relegate the discussion of roots of auxiliary Y-functions and the formulation of the TBA and exact Bethe equations
to appendices \ref{app:Yfunctions}  and \ref{app:TBA}. There we also briefly comment on how to obtain the TBA system for $L\geq8$.

Here, instead, we focus our attention on some peculiar properties of $Y_Q$ functions for states with complex rapidities, which were also found in \cite{Arutyunov:2011mk}. A crucial observation there is that  depending on the location of the rapidities on the $z$-torus some $Y_Q$-functions may have poles inside the analyticity strip. As a result, there is a root of $1+Y_Q$ located in the vicinity of a pole. If the rapidities lie just outside the analyticity strip, this leads to the appearance of extra terms in the 
 TBA equations as well as the dispersion relation and total momentum quantization condition. 

This is precisely what happens in the case $L=6$ for $Y_2$. Let us indicate from now on the rapidities of the magnons as $u_i^{(1)}$. They obey the exact Bethe equations
$1+Y_{1*}(u_i^{(1)})=0$.
Since we have for $Y_2$ that
\bea
\la{Y2poles}
Y_2(u_2^{(1)+})=\infty\,,\quad\quad Y_2(u_3^{(1)-})=\infty\,,
\eea
and $u_2^{(1)+}$ and $u_3^{(1)-}$ are
close to the real line then  there exist two complex conjugate roots $u^{(2)}_{2,3}$ close to $u^{(1)}_{2,3}$ such that
\bea
\label{Y2zeros}
1+Y_2(u_2^{(2)+})=0\,\quad\quad1+Y_2(u_3^{(2)-})=0\,.
\eea
Similar relations can be written also for $Y_3$ close to $u_2^{(1)++}$, but as it turns out, in the case of rapidities just outside the physical strip we can cast the TBA equations in a form that depends only on the usual roots $u_{2,3}^{(1)}$ and the (shifted) roots $u_{2,3}^{(2)}$.

Taking  {\it e.g.} the first equality in \eqref{Y2zeros} and expanding $Y_2(u)=\frac{{\rm Res}\, Y_2(u)}{u-u_2^{(1)+}}$  around the pole at $u_2^{(1)+}$, one gets
\bea
\la{deltauexpansion}
-\big(u_2^{(2)}-u_2^{(1)}\big)={\rm Res}\, Y_2\big(u_2^{(1)}\big)+\frac{\partial\,{\rm Res}\, Y_2}{\partial u}\big(u_2^{(1)+}\big)\,(u_2^{(2)}-u_2^{(1)})+\dots\, .
\eea
For small residue of $Y_2$  this relation implies  that $u_2^{(2)}-u_2^{(1)}$ is of order of ${\rm Res}\, Y_2$ which for small $g$ is $g^{2L}$. 
It is also worth noticing that due to the presence of the poles (\ref{Y2poles}) which are very close to the real line and almost pinch it, $Y_2(u)$ will take large values around $u={\rm Re}\big(u_2^{(1)}\big)$.

\subsection{Wrapping corrections for $L=6$ at $\cO(g^{12})$}
We are interested in the first correction to the energy, which can be found from a perturbative expansion of the energy formula \cite{Arutyunov:2011mk}
\bea
\la{energyTBA}
E&=&J+\sum_{i=1}^3\mathcal{E}(u_i^{(1)})
-{1\ov 2\pi}\sum_{Q=1}^{\infty}\int_{-\infty}^\infty\, du {d\tilde{p}_Q\ov du}\log(1+Y_Q)\\
&&\quad -i\tilde{p}_2(u_2^{(1)+})+i\tilde{p}_2(u_2^{(2)+})
-i\tilde{p}_2(u_3^{(2)-})+i\tilde{p}_2(u_3^{(1)-})\,,
\nonumber
\eea
where we used the fact that for $L=6$ the rapidities lie just outside the analyticity strip. 
 To compute $\mathcal{E}(u_i^{(1)})$ to the order $g^{12}$, it is sufficient to consider the asymptotic expression of the rapidities found by solving (\ref{allloopBYtwist}) and one obviously reproduces (\ref{divergenEn}) from the first two terms in \eqref{energyTBA} since they correspond to (\ref{allloopdispersion}). 

The leading perturbative correction due to wrapping effects can be found by expanding  the remaining terms,
\bea
\label{Ewrap}
\Delta E^{\rm(wrap)}&=&-\frac{1}{2\pi}\sum_{Q=1}^{\infty}\int_{-\infty}^\infty\, du\, {d\tilde{p}_Q\ov du}\,Y^\circ_Q\\
&&-i\frac{\partial \tilde{p}_2}{\partial u}(u_2^{(1)+})\,{\rm Res}Y_2^\circ\big(u_2^{(1)+}\big)+i\frac{\partial \tilde{p}_2}{\partial u}(u_3^{(1)-})\,{\rm Res}Y_2^\circ\big(u_3^{(1)-}\big)\,,
\nonumber
\eea
where we made use of (\ref{deltauexpansion}) and replaced everywhere $Y_Q$ by its asymptotic expression $Y_Q^\circ$, which can be found in appendix \ref{app:Yfunctions}. Furthermore, at this order only the one-loop rapidities $u_i^{(1)}$ are needed.

The final result is similar to the correction one would na\"ively expect from L\"uscher's formula, with the important addition of the terms in the second line which are dictated by the contour deformation trick. It is worth noticing that, since $Y_Q^\circ(u)\geq0$, the contribution of the first line alone is negative and for this reason can never cancel the small $\phi$ divergence in (\ref{divergenEn}).

In the case $L=6$ the computation of $\Delta E^{\rm(wrap)}$ can be readily performed. As discussed above, the separation between the poles of $Y_2^\circ$ at $u_2^{(1)+}$ and $u_3^{(1)-}$ vanishes as $\phi^6$ for small $g$, as indicated by (\ref{onelooptwistedsol}). Thus, the contributions divergent in the limit $\phi\to0$ come from the integral of $Y^\circ_2$ and from the residues on the second line of \eqref{Ewrap}. Computing $\Delta E^{\rm(wrap)}$ and adding it to the asymptotic contribution, one finds that \emph{all divergent terms cancel out}, giving in the limit $\phi\to0$ the following result
\bea
\la{energyg12finite}
E=6 &+& 3 g^2 - \frac{9}{4} g^4 + \frac{63}{16} g^6 - \frac{621}{64} g^8 - 
 \frac{9}{256}(8\zeta(3)-783 ) g^{10}\\
&+&\left(-\frac{567}{128}\zeta(9)+\frac{189}{64}\zeta(5)+\frac{243}{128}\zeta(3)-\frac{84753}{1024} \right)g^{12}+\cO(g^{14},\phi)\,.
\nonumber
\eea
The cancellation of the divergencies would not be possible without  the terms involving $u_{2,3}^{(2)}$.
This provides the first justification of the energy formula (\ref{energyTBA}) which does not rely on the contour deformation trick.

\subsection{Comments on the $g^{14}$ correction}
The cancellation of the divergencies at $g^{12}$  indicates that, when wrapping corrections are properly accounted for, the energy should not suffer  from any singularity even at higher loop orders. On the other hand, considering the solution of the asymptotic Bethe ansatz (\ref{allloopBYtwist}), we find that not only the energy at $g^{14}$ but also the rapidities at $g^{12}$ \emph{are divergent}  when the twist is removed. The mirror TBA is expected to render  at least the energy formula finite.

Unfortunately, even for the simplest $L=6$ state, computing exactly the wrapping correction to the energy at order $g^{14}$ is a non-trivial task, conceptually similar to finding the five-loop energy of the Konishi multiplet  \cite{AFS10,BH10a}, but much more involved because of the sophisticated analytic structure of the TBA system under consideration.

To progress with the calculation of the energy at $g^{14}$, one needs to know the rapidities $u_i^{(1)}$ at six loops. These cannot be found just by solving the Bethe-Yang equations: one has to consider the exact Bethe equations
\bea
\log Y_{1*}(u_i^{(1)})=(2n+1)\pi\,i\,,\quad\quad n\in\mathbb{Z}\, .
\eea
These equations are spelled out explicitly in appendix \ref{app:TBA} and they involve auxiliary Y-functions as well as their roots. In a perturbative expansion, the exact Bethe equations can be written  as
\bea
0=\log Y_{1*}(u_i^{(1)})-(2n+1)\pi\,i=\log{\rm BY}^{(i)}(u_i^{(1)})+\delta \mathcal{R}^{(i)}(u_i^{(1)})\, ,
\eea
where ${\rm BY}^{(i)}$ represents the  Bethe-Yang contribution for particle $i$ and $\delta \mathcal{R}^{(i)}$ is a correction of order $g^{12}$ (which also depends on the other rapidities, auxiliary Y-functions and roots). If we expand the rapidities $u_i^{(1)}$ around the asymptotic solution $u_i^\circ$,
\bea
u_i^{(1)}=u_i^\circ+\delta u_i^{(1)}\,,
\eea
we find that the exact Bethe equations can be rewritten as
\bea
\la{expandedEBE}
0=\sum_{k=1}^3\frac{\partial{\rm BY}^{(i)}}{\partial u_k^\circ}(u_i^\circ)\,\delta u_k^{(1)}+\delta \mathcal{R}^{(i)}(u_i^\circ)+\cO(g^{14})\,,
\eea
where we used that by construction ${\rm BY}^{(i)}(u_i^\circ)=1$.

These three coupled equations are supplemented by the quantization condition of the total momentum $P=2\pi m+M\phi/L$, where the total momentum is given by
\begin{align}
P=&
\sum_i\, p_i-{1\ov 2\pi}\int_{-\infty}^\infty\, du\, {d\tilde{\E}_Q\ov du}\log(1+Y_Q)
\nonumber\\
&-i\tilde{\E}_2(u_2^{(1)+})+i\tilde{\E}_2(u_2^{(2)+})
-i\tilde{\E}_2(u_3^{(2)-})+i\tilde{\E}_2(u_3^{(1)-})\,.
\la{momen2}
\end{align}
Notice that the quantization condition is non-trivial because unlike most other cases, {\it e.g.} that of the Konishi operator where the rapidities come in pairs of opposite signs, $P$ cannot be immediately seen to vanish due to the parity properties of $Y_Q$-functions.

A natural question one may ask is whether the wrapping corrections to rapidities $\delta u_i^{(1)}$ eliminate the divergent contributions in the asymptotic result at $g^{12}$. In that case, it should be
\bea
\la{deltauguess}
\delta u_i^{(1)}=-\left({\rm divergent\ part\ of\ } u_i^\circ\right) +O(\phi^0)\,.
\eea  
Without having to solve the complicated set of equations (\ref{expandedEBE}), we can plug our guess (\ref{deltauguess}) into the total momentum quantization condition and check whether it is satisfied. The advantage of this strategy is that at the order $g^{12}$ we  can expand  (\ref{momen2}) as 
\begin{align}
P=&
\sum_i\, p(u_i^\circ)+\sum_i\, \frac{\partial{p}}{\partial u}(u_i^\circ)\,\delta u_i^{(1)}-{1\ov 2\pi}\int_{-\infty}^\infty\, du\, {d\tilde{\E}_Q\ov du}\,Y_Q^\circ
\nonumber\\
&-i\frac{\partial\tilde{\E}_2}{\partial u}(u_2^{\circ\,+})\,{\rm Res}\, Y_2(u_2^{\circ\,+})+i\frac{\partial\tilde{\E}_2}{\partial u}(u_3^{\circ\,-})\,{\rm Res}\, Y_2(u_3^{\circ\,-})+\cO(g^{14})\,.
\la{momen3}
\end{align}
where the only non-asymptotic objects appearing in (\ref{momen2}) are precisely $\delta u_i^{(1)}$.

Surprisingly, we find that the guess (\ref{deltauguess}) is incompatible with the total momentum condition; in fact it would make $P$ divergent as $\phi\to0$. This implies that \emph{the individual rapidities found from exact Bethe equations remain divergent in perturbative theory}.

The only way of checking whether the $g^{14}$ wrapping correction to the energy makes it finite for small $\phi$ is to deal with the full set of TBA equations and expand them around the asymptotic solution and then in powers of $g$. This is straightforward but cumbersome, and is done in appendix \ref{app:linearizedTBA} for the case $L=6$. The linearized TBA system ends up to be more complicated than in the case of the Konishi operator. In particular, the linearized system for the correction to $Y_{M|vw}$-functions does not decouple from the other auxiliary equations, which makes it hard to find an analytic solution.

On the other hand, if we focus on the most $\phi$-divergent part of the corrections to rapidities (which in turn determine the most divergent part of the corrections to the energy) it is relatively easy to see that once again \emph{the wrapping effects precisely cancel the asymptotic divergence}. The compatibility of this cancellation with (\ref{momen2}) can also be seen as a non-trivial check of the formula for the total momentum.

In conclusion, we find strong evidence of a general mechanism by which the TBA description of the exceptional operators  can be obtained by introducing a twist $\phi$ as a regulator. Even if the TBA system can be found from the asymptotic data only when $g\lesssim \phi$, and therefore never, strictly speaking, at $\phi=0$, the resulting physical predictions will be regular in $\phi$ when wrapping effects are accounted for. Therefore, we can compute the perturbative energy for small $\phi$ and then take the limit $\phi\to0$ in the final result. 

Even if in principle a similar strategy could be repeated to find energies at finite $g$, this would require to (numerically) solve the full TBA system for several values of $\phi$ in order to extrapolate to $\phi\to0$ result. This would be practically unfeasible, and it is therefore important to look for an alternative TBA description of these operators, which does not resort to introducing a regulator. 

\section{TBA with exceptional rapidities}

The twisted TBA approach provides a way to compute the anomalous dimensions of exceptional operators in perturbative gauge theory. However, it leaves open a question of determining the dimensions at any value of the coupling constant. In this section we propose a set of  TBA equations which allows one to calculate the dimensions of these operators at any value of $\lambda$. 

The main idea is that since an exceptional operator is dual to a string theory state which is composed of a fundamental particle and a two-particle bound state  with maximum allowed momenta $\pm \pi$, the Bethe roots in the gauge theory normalization for any exceptional state  are in fact independent of the coupling constant:
 $u_1=0\,,u_2=-i\,,u_3=i$. The roots $u_{2,3}$ satisfy the bound state condition, and since their real part is 0, they are on the cuts of $x_s^\pm$ functions. According to \cite{AF07}, they must lie on the same sides of the cuts, and therefore, we propose that  
the exact Bethe rapidities (in the string theory normalization which will be convenient to write the TBA equations in this section) for any exceptional state are equal to 
\be\la{exactu}
u_1=+{i 0\ov 2}\,,\quad u_2 = -{i\ov g}-i0\,,\quad u_3={i\ov g}-i0\,.
\ee
With this choice of the signs in front of $i0$, the fundamental particle and the bound state composed of $u_{2,3}$ have momenta $+\pi$ and $-\pi$ respectively, if one uses Mathematica's conventions for branch cuts.
Then the root $u_2$ lies in the intersection of the mirror and string regions, and $u_3$ is in the intersection of the string  and the second mirror regions. Notice that it is different from the state analyzed in \cite{Arutyunov:2011mk} where  the rapidity $u_3$ was in the intersection of the string and the anti-mirror regions. The location of the rapidities on the $z$-torus is shown on figure \ref{trus_rapidities}, and in terms of the $z$-rapidity variable all Y-functions and dispersion relations are meromorphic in the vicinities of these points.
\begin{table}
\begin{center}
\begin{tabular}{|c|c|c|}
\hline Y${}^o$-function  & Zeroes   & Poles\\
\hline $Y_{M|w}$   & $0^2$ &  \\
\hline $1+Y_{M|w}$ & $-i/g\,,\ +i/g$ & $-(M+2)i/g\,,\ (M+2)i/g$  \\
\hline $Y_{1|vw}$ &$0^2$ & \\
\hline $1+Y_{M|vw}$ && $Mi/g\,,\ -Mi/g$ \\
\hline $Y_-$      & $-2i/g\,,\ 2i/g$ & $0^2$ \\
\hline $Y_+$       & & $0^2\,,\ -i/g$\\
\hline $1-Y_-$   & $-i/g\,,\ i/g$ & \\
\hline $1-Y_+$   &&\\
\hline $Y_1$   & $0^2$&$-i/g\,,\ +i/g$ \\
\hline $Y_2$   & &$0^2$ \\
\hline $Y_Q\,, Q\ge 3$   & &$i(Q-2)/g\,,\ -i(Q-2)/g$ \\
\hline
\end{tabular}
\end{center}
\caption{Relevant roots and poles of asymptotic  Y-functions within the mirror region. $0^2$ means either a double zero or a double pole at $0$.
}
\label{Yfun}
\end{table}

These rapidities lead to a quite simple analytic structure of asymptotic Y-functions with double poles and zeroes at the origin of the mirror $u$-plane, see Table \ref{Yfun}, and it is natural to assume that the exact Y-functions would have the same analytic properties.\footnote{Let us mention that Y-functions with double poles and zeroes at the origin are typical for boundary TBA, see {\it e.g.}   \cite{Bajnok:2007ep,Correa:2012hh,Drukker:2012de}.}

\subsection{TBA equations}
In this subsection we list the simplified and hybrid TBA equations for the exceptional states. 
They can be obtained from the ones discussed in \cite{Arutyunov:2011mk} by sending the roots $r_M$ to 0, and $u_i^{(1)}\,, u_i^{(2)}$ to $u_i$. 
The only exception is the hybrid equations for $Y_Q$ where one
should take care of the fact that the root $u_3$ is located in the intersection of the string region and the second mirror region but not in  the anti-mirror region as it was in \cite{Arutyunov:2011mk}. The TBA equations below are consistent with the analytic structure of  Y-functions in Table \ref{Yfun} supplemented by the conditions $Y_{1_*}(0)=Y_{1_*}(-i/g)=Y_{1_*}(i/g)=-1$.

\subsubsection*{Simplified equations for  $Y_{M|w}$}

\be
\log Y_{M|w} =2\log S({i\ov g} + v)+  \log(1 +  Y_{M-1|w})(1 +
Y_{M+1|w})\star s+ \delta_{M1}\, \log{1-{1\ov Y_-}\ov 1-{1\ov Y_+} }\hstar s \,.~~~~~
\ee

\subsubsection*{Simplified equations for  $Y_{M|vw}$}
\begin{align}
\log Y_{M|vw} = &  2\delta_{M1}\log S({i\ov g} + v)  +
\log(1 +  Y_{M-1|vw} )(1 +  Y_{M+1|vw})\star
s\nonumber\\
&  + \delta_{M1} \log{1-Y_-\ov 1-Y_+}\hstar s- \log(1 +  Y_{M+1})\star s\,.
\end{align}

\subsubsection*{Simplified equations for  $Y_\pm$}
\begin{align}\la{seqypovm}
\log {Y_+\ov Y_-} &= \,   \log(1 +  Y_{Q})\star K_{Qy} - \sum_i \log S_{1_*y}(u_i,v)\,,\\ \nonumber
 \log {Y_+ Y_-} &=\  2\log {1+Y_{1|vw}
 \ov 1+Y_{1|w}}\star s- \log\left(1+Y_Q
\right)\star K_Q
+2\log(1 + Y_{Q}) \star K_{xv}^{Q1}\star s
 \\\la{seqypm}
&\qquad\qquad\quad-4\log S({i\ov g} + v)-\sum_i \log {S_{xv}^{1_*1}(u_i,v)^2\ov S_2(u_i- v)}\star s 
\,.
\end{align}
It is worth mentioning that since the driving terms in the equations above satisfy the discrete Laplace equation 
$$\S_Q(v-{i\ov g})\S_Q(v+{i\ov g})= \S_{Q-1}(v)\S_{Q+1}(v)\,,\quad \S_0(v)=1\,,$$
they can be written as
\begin{align} &- \sum_i \log S_{1_*y}(u_i,v)= -\log S_{1_*y}(0,v)-\log S_{2_*y}(0,v)\,,\\
  \nonumber
&-\sum_i \log {S_{xv}^{1_*1}(u_i,v)^2\ov S_2(u_i- v)}\star s =-\log {S_{xv}^{1_*1}(0,v)^2\ov S_2(0- v)}\star s-2\log S_{xv}^{2_*1}(0,v)\star s  +\log S_2(0- v)
\,.
\end{align}
This shows that  the driving terms in eqs.(\ref{seqypovm},\ref{seqypm}) can be understood as appearing
not  because of the zeroes of $1+Y_{1_*}$ at $u=0\,,-i/g\,, i/g$ in the string $u$-plane but due to the zeroes of $1+Y_{1_*}$ and $1+Y_{2_*}$ at $u=0$ in the string $u$-plane.
It is consistent with the interpretation of  an  exceptional state as a bound state of a fundamental particle and a two-particle bound state with rapidities equal to 0. This interpretation however requires using integration contours different from the ones described in \cite{Arutyunov:2011mk}.

\subsubsection*{Simplified TBA equations for $Y_Q$}
\bigskip
 \noindent
$\bullet$ $\ Q\ge 3\ $
\bea
\log Y_{Q}&=&\log{\left(1 +  {1\ov Y_{Q-1|vw}} \right)^2\ov (1 +  {1\ov Y_{Q-1} })(1 +  {1\ov Y_{Q+1} }) }\star s
\,.~~~~~~~
\eea
 \noindent
$\bullet$ $\ Q=2\ $
\bea
\log Y_{2}&=&-2\log S({i\ov g}-v)+\log{\left(1 +  {1\ov Y_{1|vw}} \right)^2\ov (1 +  {1\ov Y_{1} })(1 +  {1\ov Y_{3} }) }\star_{p.v} s
\,.~~~~~~~
\eea

\subsubsection*{Hybrid TBA equations for $Y_Q$}
To make the presentation transparent, we introduce a function which combines the terms on the right hand side of the hybrid ground state TBA equation ($L_{\rm TBA}=J+2$)
\bea
G_Q(v)&=& - L_{\rm TBA}\, \tH_{Q} +\log \left(1+Y_{Q'} \right)
\star (K_{\sl(2)}^{Q'Q}+2 s\star K_{vwx}^{Q'-1,Q})\\
&+&  2 \log \(1 + Y_{1|vw}\) \star s \hstar K_{yQ} +2\log(1+Y_{Q-1|vw})\star s\nonumber \\
&  -&  2  \log{1-Y_-\ov 1-Y_+} \hstar s \star K^{1Q}_{vwx} +  \log
{1- \frac{1}{Y_-} \ov 1-\frac{1}{Y_+} } \hstar K_{Q}  +  \log
\big(1-\frac{1}{Y_-}\big)\big( 1 - \frac{1}{Y_+} \big) \hstar
K_{yQ} \, . \nonumber  \eea
With the help of $G_Q$, the hybrid TBA equations for $Y_Q$ read as
\begin{align}\la{TBAYQ}
\log Y_Q(v) &= G_Q(v) -\sum_i\log S_{\sl(2)}^{1_*Q}(u_i,v)+4\log S\star_{p.v.} K_{vwx}^{1Q}(-{i\ov g},v) \\ \nonumber
& -\log S_Q(-{i\ov g}-v)S_{yQ}(-{i\ov g},v)S_Q(-v)S_{yQ}(0,v)S_Q({2i\ov g}-v)S_{yQ}({2i\ov g},v)
 \, . \end{align}

It is important to stress that since the  location of  the Bethe rapidities is exactly known the only parameters in the TBA equations for exceptional operators are the charge $J$ (or equivalently the operator length $L=J+3$) and  the coupling constant $g$. In this respect these TBA equations are of the same level of complexity as the ones for the ground state of any integrable model.

\subsection{Exact Bethe equations}

To construct the TBA equations by using the contour deformation trick one has to assume that $1+Y_{1_*}$ has zeroes at $u=0\,,-i/g\,, i/g$ in the string plane. On the other hand once the equations have been derived one can use the analytic continuation to calculate $Y_1$ at these points. Thus,  the conditions
\be
Y_{1_*}(0)=-1\,,\quad Y_{1_*}(-{i\ov g})=-1\,,\quad Y_{1_*}({i\ov g})=-1\,,
\ee
on $Y_{1_*}$ must follow from the 
 TBA equations. This imposes nontrivial consistency conditions on the TBA equations which we discuss in this subsection.

\subsubsection*{Bethe equation at $u_1=0$: $Y_{1_*}(0)=-1$}

We begin by  showing  that $Y_{1_*}(0)=-1$. Indeed analytically continuing the equation for $Y_1$ to real $v$ one gets
\begin{align}
\log Y_{1_*}(v) &= G_{1_*}(v) -\sum_i\log S_{\sl(2)}^{1_*1_*}(u_i,v)\nonumber \\ \nonumber&
+4\log {\rm Res\ } S\star K_{vwx}^{11_*}(-{i\ov g},v) +2\log S_{vwx}^{11_*}(-{i\ov g},v) - 4 \log(-v-{2i\ov g}){x_s^-(0) -{1\ov x_s^-(v)}\ov x_s^-(0)-{1\ov x_s^+(v)}}
\nonumber \\ \nonumber
& -\log  S_1(-{i\ov g}-v)S_{y1_*}(-{i\ov g},v)S_1(-v)S_{y1_*}(0,v)S_1({2i\ov g}-v)S_{y1_*}({2i\ov g},v)
 \, . \end{align}
 Then one finds that the imaginary part of $G_{1_*}(v)$
 in the limit $v\to 0$ is equal to $i\pi (J+2)$ because all the kernels in $G_{1_*}(v)$ are antisymmetric at $v=0$, and the real part of $G_{1_*}(v)$ is given by the usual expression 
\begin{align}
 {\rm Re}\ G_{1_*}(v)= -\sum_i \log S_{1_*y}(u_i,v) \cstar K_1
  \, . \end{align}
One can then easily check that in the limit $v\to 0$
\begin{align}
& -\sum_i \log S_{1_*y}(u_i,v) \cstar K_1 -\sum_i\log S_{\sl(2)}^{1_*1_*}(u_i,v)\nonumber \\ \nonumber&
+4\log {\rm Res\ } S\star K_{vwx}^{11_*}(-{i\ov g},v) +2\log S_{vwx}^{11_*}(-{i\ov g},v) - 4 \log(-v-{2i\ov g}){x_s^-(0) -{1\ov x_s^-(v)}\ov x_s^-(0)-{1\ov x_s^+(v)}}
\nonumber \\ \nonumber
& -\log  S_1(-{i\ov g}-v)S_{y1_*}(-{i\ov g},v)S_1(-v)S_{y1_*}(0,v)S_1({2i\ov g}-v)S_{y1_*}({2i\ov g},v) = 0 \ \ {\rm mod}\ 2\pi i 
 \, , \end{align}  
 and therefore 
  \begin{align}
\log Y_{1_*}(0) = i\pi (J+2)
 \, . \end{align}
 Thus if $J$ is odd as it is for exceptional operators then $Y_{1_*}(0)=-1$.

 \subsubsection*{Bethe equation at $u_2=-i/g$: $Y_{1}(-i/g)=-1$}
 
 To show that $Y_{1_*}(-i/g)=-1$ we  notice that $u_2=-i/g-i0$ is in the mirror region, and therefore $Y_{1_*}(u_2)=Y_{1}(u_2)$. Moreover, since
 we approach $-i/g$ from the mirror real line, we can always use the mirror-mirror kernels in \eqref{TBAYQ}. 
 Then to show that $Y_{1}(-i/g)=-1$ we use that all Y-functions are even, and all the kernels in \eqref{TBAYQ} satisfy 
\be
K(t,v) = K(-t,-v)\,,
\ee
 and  therefore for any even function $f$
 \be
 2f\star K(v) = f\star K(v) + f\star K(-v)\equiv f\star \big(K(v)+K(-v)\big)\,.
 \ee
 Thus we have the following equality
 \begin{align}\la{TBAYQ1}
2\log Y_1(v) &= G_1(v)+G_1(-v) -2\sum_i\log S_{\sl(2)}^{1_*1}(u_i,v)\\ \nonumber
&+4\log S\star_{p.v.} \big( K_{vwx}^{11}(-{i\ov g},v) +K_{vwx}^{11}(-{i\ov g},-v)\big) \\ \nonumber
& -2\log S_1(-{i\ov g}-v)S_{y1}(-{i\ov g},v)S_1(-v)S_{y1}(0,v)S_1({2i\ov g}-v)S_{y1}({2i\ov g},v)
 \, . \end{align}
Now we want to take the limit $v\to -i/g$. Since all the kernels satisfy the discrete Laplace equation we would na\"ively get 
\be\la{G112}
G_1(-i/g)+G_1(i/g) =G_2(0) - \log \left(1+Y_{2}(0) \right)\,,
\ee
 where the last term appears because of the pole in $K_{\sl(2)}^{21}(t,v)$ at $t=\pm i/g$. The kernel $K_{yQ}$ also has a pole there and it produces the term $2 \log \(1 + Y_{1|vw}\) \star s$ which is in $G_2$, and it could produce the term $\log
\big(1-\frac{1}{Y_-}\big)\big( 1 - \frac{1}{Y_+} \big)$ but it vanishes because $Y_\pm(0)=-\infty$. 
The only problem with \eqref{G112} is that $Y_{2}(0)=\infty$, and therefore we should deal with the term $F_1\equiv -\log \left(1+Y_{2}\right)\star K_1$ more carefully. 
 We represent it in the form
 \be
F_1(v)= -\int\, dt\, \log{1+Y_{2}(t)\ov 1+{C^2\ov t^2}} K_1(t-v) - \int\, dt\, \log\big(1+{C^2\ov t^2}\big) K_1(t-v)\,,
 \ee
 where $C^2=\lim_{t\to 0}\, t^2Y_{2}(t)$. 
 The first term then represents no problem and one gets 
 \begin{align}
2F_1(\eps -{i\ov g})=& -\int\, dt\, \log\big(1+Y_{2}(t)\big) K_2(t-\eps) \\ \nonumber
& +\int\, dt\, \log\big(1+{C^2\ov t^2}\big) K_2(t-\eps) - 2\int\, dt\, \log\big(1+{C^2\ov t^2}\big) K_1(t-\eps +{i\ov g})\,,
 \end{align}
 where $\eps$ is infinitesimally close to 0 with positive imaginary part.
 The integral on the second line can be computed, and expanding it in powers of $\eps$ one gets
 \be
 2F_1(\eps-{i\ov g})= -\log\big(1+Y_{2}\big) \star K_2(0) -\log{C^2\ov \eps^2}-i\pi\,.
 \ee
 Thus, the formula \eqref{G112} contains the extra $i\pi$ term, and takes the form
\be\la{G112b}
G_1(-i/g)+G_1(i/g) =G_2(\eps) - \log Y_{2}(\eps)-i\pi +o(\eps)\,.
\ee
Taking into account the TBA equation for $Y_2$ one gets 
 \begin{align}\la{TBAYQ1b}
2\log Y_1(v) &= -i\pi+\sum_i\log S_{\sl(2)}^{1_*2}(u_i,v+{i\ov g}) -2\sum_i\log S_{\sl(2)}^{1_*1}(u_i,v)
 \\ \nonumber
& -2\log S_1(-{i\ov g}-v)S_{y1}(-{i\ov g},v)S_1(-v)S_{y1}(0,v)S_1({2i\ov g}-v)S_{y1}({2i\ov g},v)
 \\ \nonumber
& +\log S_2(-{i\ov g}-v)S_{y2}(-{i\ov g},v)S_2(-v)S_{y2}(0,v)S_2({2i\ov g}-v)S_{y2}({2i\ov g},v)
 \, , \end{align}
where $v=\eps-{i\ov g}$. Taking the limit $\eps\to 0$ one finally gets
\be
\log Y_1(-{i\ov g}) = -i\pi\,.
\ee

In the same way one can show that $Y_1({i\ov g}) = -1$ (or one can use the Y-system equation for $Y_1$), and then the condition $Y_{1_*}({i\ov g}) = -1$ can be proven by using  the crossing symmetry relations as was done in \cite{Arutyunov:2011mk}. Let us finally mention that it should be possible to show that the TBA equations imply in addition $Y_{2_*}(0)=-1$ because the 
particles with rapidities $\pm i/g$ can be thought of as constituents of a two-particle bound state with rapidity equal to 0. This however requires a careful analytic continuation of  the hybrid TBA equation for $Y_2$ to the string $u$-plane through the cut at $-2i/g$, and we will not pursue this here. 
 
 \subsubsection*{Scaling dimensions of exceptional operators}
 
Scaling dimensions  of exceptional operators or energies of dual string states are found from the usual formula 
 \begin{align}
\Delta-J=E-J=&\
\sum_{i}\E(u_i)
-{1\ov 2\pi}\int_{-\infty}^\infty\, du {d\tilde{p}_Q\ov du}\log(1+Y_Q)
\nonumber\\
&\quad =\sqrt{1+4g^2}+\sqrt{4+4g^2}-{1\ov 2\pi}\int_{-\infty}^\infty\, du {d\tilde{p}_Q\ov du}\log(1+Y_Q)\,,
\la{Ener1}
\end{align}
where we used the exceptional rapidities of the particles. This formula shows that at large $g$ the first two terms in \eqref{Ener1} which come from the dispersion relation are proportional to $g$. On the other hand for finite $J$ and large $g$ the scaling dimension of these operators should behave as $\sqrt g$. Thus, the linear term should be canceled by the contribution coming from the $Y_Q$-functions.  This is different from the expected large $g$ behaviour of  two-particle states  studied in \cite{AFS09,Frolov:2012zv}.  It would be interesting to understand if the linear term comes entirely from the pole contribution of  $Y_2$.

\subsection{Leading TBA correction up to $g^{10}$}

The proposed TBA equations are based on the assumption that the rapidities of exceptional states are given exactly by \eqref{exactu}. These rapidities are obviously very different from the rapidities of the states in the twisted theory which diverge in the limit $\phi\to 0$ at least in the perturbation theory.
Still, the TBA equations should produce the same perturbative expansion of the scaling dimensions of  exceptional operators as the one we obtained from the twisted TBA equations in the previous section. In this and next subsections we compute the scaling dimension of  the shortest exceptional operator of length $L=6$ and show that it coincides with the twisted TBA result.  We will use the gauge theory normalization of  rapidities in which the exact Bethe roots are $0,\pm i$.

Let us recall that the finite-size corrections to the energy of  the twisted exceptional operator for finite $\phi$ start exactly at $g^{12}$ as expected for an operator  of length $L=6$ from the $\su(2)$ sector.  Thus up to $g^{10}$ one can just use the dispersion relation and the BY equations. Then, as was shown in the previous section, one gets
\be\la{en1}
E_{\phi=0}=6+3 g^2-\frac{9 g^4}{4}+\frac{63
   g^6}{16}-\frac{621 g^8}{64}-\frac{9 g^{10} \zeta (3)}{32}+\frac{7047
   g^{10}}{256}\,.
\ee
On the other hand if one uses the energy formula \eqref{Ener1} with  the exceptional Bethe roots,  then the contribution coming from the dispersion relation 
is just given by the first two terms and its expansion up to $g^{10}$ produces
\be\la{enas}
E^{\rm asym}=\sqrt{1+4g^2}+\sqrt{4+4g^2} \approx 6+3 g^2-\frac{9 g^4}{4}+\frac{33
   g^6}{8}-\frac{645 g^8}{64}+\frac{3591 g^{10}}{128}\,.
\ee
The two formulas obviously become different already at the $g^{6}$ order.
Thus the  finite-size corrections in the case of  the TBA with exceptional rapidities must appear at the $g^{6}$ order which from the field theory point of view is half-wrapping. 
We know that perturbative expansion of all $Y_Q$-functions begins at $g^{12}$ and therefore any $Y_Q$-function regular on the real line begins to contribute to the energy at the $g^{12}$ order. The only exception is 
$Y_2$-function which has a double pole at zero (if $\phi=0$).  As a result the perturbative expansion of the integral $\int du {d\tilde p\ov du} \log(1+Y_2)$  starts at the $g^{6}$ order.  Thus, up to the $g^{10}$ order one should get the same energy \eqref{en1} by keeping only $Y_2$ in TBA equations and the energy formula. Therefore, the formula of interest up to $g^{10}$ is 
\be\la{en2}
E =E^{\rm asym} - {1\ov 2\pi}\int du {d\tilde p_2\ov du} \log(1+Y_2)
\ee
where $E^{\rm asym}$ is given by 
\eqref{enas}.
Up to the $g^{10}$ order we  only need the coefficient of the double pole at $u=0$ up to the $g^{16}$ order
\bea\nonumber
Y_2(u)&=& \la{Y2p}
\frac{9 g^{12} \left(3 g^4 (8 \zeta (3)+15)-24
   g^2+8\right)}{2048 u^2} + const +\cO(u^2)\,.
\eea 
Then computing the integral in \eqref{en2} one finds
\be\la{enLues}
E^{\rm pole}=- {1\ov 2\pi}\int dv {d\tilde p_2\ov dv} \log(1+Y_2)=-\frac{9 g^{10} \zeta (3)}{32}-\frac{135
   g^{10}}{256}+\frac{3 g^8}{8}-\frac{3 g^6}{16}\,,
\ee
where in $Y_2$ we  only kept the $1/u^2$ term. 

Adding \eqref{enLues} to \eqref{enas}, one gets
precisely  \eqref{en1}.

\subsection{Next-to-leading TBA correction at $g^{12}$}

The agreement between the energies observed in the previous subsection  should also hold at the $g^{12}$ order where one should calculate the usual contributions from all $Y_Q$-functions. In addition one also has to take into account the TBA correction to the coefficient of the double pole of $Y_2$ which is of the $g^{18}$ order.  

\subsubsection*{Linearization of the TBA equations}

It is well-known that  at small $g$ Y-functions get TBA corrections beyond their asymptotic form $Y^\circ$.  Computing the leading TBA  corrections requires  linearization of the TBA equations which can be done by representing  any Y-function as follows
\begin{equation}
Y(u)=Y^\circ(u)\,\Big(1+\mathscr{Y}(u)\Big)\,.
\end{equation}
Since the Bethe roots do not get corrections,  the $\mathscr{Y}$'s have neither zeroes nor  poles on the real line.
Then one expands  the hybrid TBA equations  up to the first order in $\mathscr{Y}_{aux}$ while 
keeping only the contributions from 
the asymptotic $Y_Q$-functions on the r.h.s. of the equations.
It is clear that leading corrections to any $\mathscr{Y}$ are of order $g^6$ or higher, and they  come only from the pole part of $Y_2^\circ$.  
Discarding any term of $\cO(g^8)$, we find that only the following two equations  are relevant at the $g^6$ order 
\bea\la{eqY2l}
\mathscr{Y}_2&=&\log(1+Y_2^\circ)\star(K^{22}_{\sl(2)}+2s\star K^{12}_{vwx})+4\left(A_{1|vw}\,\mathscr{Y}_{1|vw}\right)\star s\,,\\
\label{linearizedYvw}
\mathscr{Y}_{M|vw}&=&A_{M-1|vw}\mathscr{Y}_{M-1|vw}\star s+A_{M+1|vw}\mathscr{Y}_{M+1|vw}\star s-\delta_{M1}\log(1+Y_2^\circ)\star s\,,~~~~~~
\eea
where we defined the coefficient
\[A_{M|vw}=\frac{Y_{M|vw}^\circ}{1+Y_{M|vw}^\circ},\quad\quad\quad M\geq1.\]
The $g^6$ contribution of  $Y_2^\circ$ to these equations can be easily computed because for any kernel $K(u,v)$ regular for real $u$ and $v$ one gets
\be
\log(1+Y_2^\circ)\star K\to R_2^\circ\,\int\,du\, \log(1+\frac{1}{u^2})\,K(R_2^\circ \, u,v)\to  {3\pi\ov 8}g^{6} \, K(0,v)\,,\ee
where $R_2^\circ$ is the square root of the coefficient of  the pole of $Y_2^\circ$
\be
Y_2^\circ = {(R_2^\circ)^2\ov u^2} + \cdots\,,\quad R_2^\circ={3\ov 16}g^{6}\big(1-\frac{3 g^2}{2}+\frac{3 g^4}{16} (8 \zeta (3)+9) +\frac{g^6}{32}  (-120 \zeta (3)-108 \zeta
   (5)-55)\big)\,.
\ee 
This also proves that 
the leading TBA corrections to Y-functions are of order $g^6$. 

\subsubsection*{Expansion of the energy formula}
Let us now assume that we know $\mathscr{Y}_2$ up to the $g^6$ order and compute the energy up to the $g^{12}$ order.
The expansion of $E^{\rm asym}$ gives
\be
E^{{\rm asym},\,(12)}=-\frac{43029 g^{12}}{512}\,.
\ee
The contribution of $Y_Q$ with $Q\neq 2$ is found from the usual formula 
\bea
E_{Y}^{(Q\neq 2)}= - {1\ov 2\pi}\sum_{Q\neq 2}\int\, du\,  Y_Q^\circ\,.
\eea
Computing the integrals and taking the sum, one obtains 
\bea\la{EnYQ13}
E_{Y}^{(Q\neq 2)}= 
g^{12} \left(\frac{135 \zeta (3)}{128}+\frac{297 \zeta
   (5)}{128}-\frac{567 \zeta
   (9)}{128}+\frac{358424597369}{580608000000}\right)
\eea
To find the contribution of $Y_2$ 
we represent the integrands in the energy formula as follows:
\bea\la{integrandY2}
\log(1+Y_2)&=&\log \frac{1+Y_2}{1+\frac{R_2^2}{u^2}} + \log\big(1+\frac{R_2^2}{u^2}\big) ,
\eea
where $R_2$ is the square root of the coefficient of  the pole of   $Y_2$ which also includes the contribution from $\mathscr{Y}_2$ and therefore can be written as 
\be
R_2=R_2^\circ\sqrt{1+\mathscr{Y}_2(0)}\,.
\ee
The first term in \eqref{integrandY2} is regular everywhere, and can be expanded in $g$ starting from $g^{12}$, and at that order depends solely on asymptotic quantities. Its contribution to the energy at the $g^{12}$ order is given by
\bea
E_{Y_2}^{\rm reg}= - {1\ov 2\pi}\int\, du\,  \log \frac{1+Y_2^\circ}{1+\frac{R_2^2}{u^2}} = \frac{15795402631}{580608000000}g^{12}\,.
\eea
The contribution of the second term  yields 
\begin{equation}
E_{Y_2}^{\rm pole}=- {1\ov 2\pi}\int\, du\, {d\tilde p_2\ov du}  \log\big(1+\frac{R_2^2}{u^2}\big)=E_{Y_2^\circ}^{\rm pole} -{3g^6\ov 32}\mathscr{Y}_2(0)+\cO(g^{14}),
\end{equation}
where $E_{Y_2^\circ}^{\rm pole}$ is the contribution due to the pole of $Y_2^\circ$
\be\la{enasII}
E_{Y_2^\circ}^{\rm pole}=\frac{3g^{12}}{256}  \big(72 \zeta (3)+54 \zeta (5)+55\big)-\frac{9 g^{10} \zeta (3)}{32}-\frac{135
   g^{10}}{256}+\frac{3 g^8}{8}-\frac{3 g^6}{16}\,.
\ee

This means that to find the energy at order $g^{12}$, we need to know the leading TBA correction to $Y_2$ at $u=0$. The correction is given by \eqref{eqY2l} which at $u=0$ can be written in the form
\bea\la{eqY2lb}
\mathscr{Y}_2(0)&=&\frac{3g^6}{32} (8\log (2)-3)+4\left(A_{1|vw}\,\mathscr{Y}_{1|vw}\right)\star s(0)\,.~~~~~~
\eea
The last term $4\left(A_{1|vw}\,\mathscr{Y}_{1|vw}\right)\star s(0)$ can be found by solving eq.\eqref{linearizedYvw} which takes the following explicit form
\bea
\label{linearizedYvwb}
\mathscr{Y}_{M|vw}(u)&=&A_{M-1|vw}\mathscr{Y}_{M-1|vw}\star s+A_{M+1|vw}\mathscr{Y}_{M+1|vw}\star s-\delta_{M1}\frac{3g^6}{8} \pi \,  s(u)\,.~~~~~~
\eea
Introducing the functions $\mathscr{X}_M(u)$ which satisfy the following difference equations
\bea
\label{linearizedX}
{\mathscr{X}_{M}(u+i)+\mathscr{X}_{M}(u-i)\ov A_{M|vw}(u)}&=&\mathscr{X}_{M-1}+\mathscr{X}_{M+1}+\delta_{M1} \,  2\pi s(u)\,,~~~~~~
\eea
the quantity $4\left(A_{1|vw}\,\mathscr{Y}_{1|vw}\right)\star s(0)$ appearing in \eqref{eqY2lb}  can be written in the form
\be
4\left(A_{1|vw}\,\mathscr{Y}_{1|vw}\right)\star s(0) = -\frac{3g^6}{4} \mathscr{X}_{1}(0)\,.
\ee
Thus summing up all the contributions one finds the energy of the exceptional state at the $g^{12}$ order 
\be
E^{(12)}= \frac{3 g^{12} (24 \mathscr{X}_{1}(0)-1512 \zeta (9)+1008 \zeta (5)+648
   \zeta (3)-28237-24 \log 2)}{1024}\,.
\ee
Comparing this formula with \eqref{energyg12finite} obtained from the twisted TBA, one gets
\be
E^{(12)}-E^{(12)}_{\phi=0}= \frac{3}{512} g^{12} (12 \mathscr{X}_{1}(0)+7-12 \log 2)\,.
\ee
Thus the two results coincide if 
\be\la{x1}
\mathscr{X}_{1}(0) = \log 2-\frac{7}{12}\approx 0.109814\,.
\ee
We could not prove this equality analytically. Solving the system \eqref{linearizedX} numerically we find that the equality \eqref{x1} holds with 
very high precision. 

\smallskip

To conclude this section let us point out that the consideration above can be easily generalized to the exceptional operator of length $L=J+3$. The $Y_2$-function begins to contribute at the $g^{L}$ order.  
The improved dressing factor contribution can be easily found at this order, and one gets that the energy of the exceptional operator  is just equal to
\be
E_{L} = J+\sqrt{1+4g^2}+\sqrt{4+4g^2} - {3\ov 2^{L-2}} g^L +\cO(g^{L+2})\,.
\ee
It is not difficult to check that at this order the same expression is obtained by using the twisted state in the limit $\phi\to 0$ \cite{BDS}.   
One can in principle go all the way till $g^{2L}$.
The only technically nontrivial part is finding the power series expansion of the dressing phase up to the $g^{L+2}$ order.

\section{Conclusions}
In this work we have provided the mirror TBA description for the exceptional class of gauge theory operators ${\cal O}_L$.
From the point of view of the Bethe Ansatz the states corresponding to these operators are singular that is the asymptotic energy 
diverges at the first wrapping order in the limit of vanishing twist. On the other hand, in the approach based on Baxter's $Q$-operator,
the same state with $M=3$ Bethe roots can be described by means of $L-M+1=L-2$ dual roots which are all regular at one loop.  
It would be interesting to see whether the dual root picture can be implemented at the level of the TBA equations. 
A natural starting point here would be to explicitly develop the all-loop Baxter equation in the $\su(2)$ sector in the spirit of  
\cite{Belitsky:2006wg}.

\smallskip

In a certain respect the operators from the family $\{{\cal O}_L\}$ are even more interesting than the Konishi operator. Indeed, the fact that their
Bethe rapidities are known exactly must simplify the numerical analysis of the corresponding TBA equations since one does not need to solve the exact Bethe equations.  Also, the rather rigid analytic structure of Y-functions  -- the presence of double poles and zeroes -- 
hints that it possibly remains the same all the way from weak to strong coupling which might help to find a proper ansatz for Y-functions at strong coupling.
This should be contrasted to the case of regular operators, where the position of zeroes and poles depends on the coupling constant 
and there are critical points \cite{AFS09,Frolov:2012zv}. 

\smallskip

Since a three-magnon state with rapidities $0,+i/g,-i/g$ can be viewed as a scattering state of a fundamental particle and a two-particle bound state with momenta $\pm\pi$, the asymptotic energy is 
$$
E^{\rm asym}=J+\sqrt{1+4g^2}+\sqrt{2^2+4g^2}\, .
$$
Therefore, at large $g$ the asymptotic energy scales as $E^{\rm asym}\sim g$. On the other hand, the operators we consider belong to the class of short operators for which the energy must scale as $\sqrt{g}\sim \sqrt[4]{\lambda}$ at strong coupling. Hence, according to the TBA description, the  contribution of $Y_Q$-functions must scale as $g$ at strong coupling and  cancel the leading term of $E^{\rm asym}$ at $g\to \infty$.
It would be interesting to verify this fact by constructing the corresponding analytic and numerical solution.

\smallskip

Let us also mention that  recently there has been an interesting development
\cite{Suzuki:2011dj}-\cite{Balog:2012zt} concerning a construction of a finite set of non-linear integral equations (NLIE), which is a complementary approach
to the TBA description of the spectrum of the $\AdS$ superstring. It would be important to see how the states corresponding to operators ${\cal O}_L$ can be accommodated within the NLIE approach. 

\smallskip

The experience we gained here with the exceptional operators brings us back to the question of the strong coupling behavior 
of a generic bound state in ${\cal N}=4$ theory discussed in \cite{Arutyunov:2011mk}.  We expect that similarly 
to what happens in the $\phi\to 0$ limit for twisted states,
the complex rapidities of a generic bound state will reach the branch points at finite value of $g$ and afterwards continue to move along the 
boundary of the string region towards 
the position of the exceptional rapidities reaching them at $g=\infty$.   To confirm this picture one has to further investigate the TBA equations 
obtained in \cite{Arutyunov:2011mk}. If true this would suggest
 a universal behavior of a generic state: when 
coupling increases eventually real rapidities move towards $-2,0,2$, while complex rapidities reach the branch points and upon passing them 
approach the exceptional rapidities. The points $-2,2$ and $0,\pm \frac{i}{g}$ would serve as attractors for all rapidities. This would classify states with a finite number of roots at strong coupling  and might explain the universal $\sqrt[4]{\lambda}$-behavior of the energy of short operators.

\section*{Acknowledgements}
We are grateful to Niklas Beisert, Nadav Drukker, Gregory Korchemsky, Matthias Staudacher and Stijn van Tongeren for useful discussion. We also thank Stijn van Tongeren for useful comments on the manuscript.
G.A. and A.S. acknowledge support by the Netherlands Organization for Scientific Research (NWO) under the VICI grant 680-47-602.
The work by G.A. is also a part of the ERC Advanced grant research programme No. 246974,  {\it ``Supersymmetry: a
 window to non-perturbative physics"}. The work of S.F. was supported in part by the Science Foundation Ireland under Grant 09/RFP/PHY2142
 and by the Institute for Advanced Studies, Jerusalem, within the Research Group Integrability and Gauge/String Theory.

\section{Appendices}
\subsection{Twisted transfer matrices and relating twist to a $\g$-deformation}
\label{app:gammadeform}
In this section we will show how the twist parameter $\phi$ that we have introduced as a mere regulator can be related to the parameters of a $\g$-deformation of $\mathcal{N}=4$ SYM. To do this let us recall that the most general $\g$-deformation imposes twisted boundary conditions on the angles $\varphi_i$ of $S^5$  as follows \cite{Frolov:2005dj}
\bea
\varphi_i(2\pi)=\varphi_i(0)-2\pi\,\epsilon_{ijk}\g_jJ_k\,,
\eea
where $\g_{j}$ are three deformation parameters, and $J_i$ are angular momenta on $S^5$ corresponding to the direction of $\varphi_i$. Let us introduce the notation
\bea
\alpha_i=-2\pi\,\epsilon_{ijk}\g_jJ_k\,,\quad
\alpha_\ell=-\frac{\a_2+\a_3}{2}\,,\quad
\alpha_r=-\frac{\a_2-\a_3}{2}\,.
\eea
The level-matching condition in the presence of such modified boundary conditions is
\bea
\la{twistedtotalmom}
P=\alpha_1+2\pi n\,,\quad n\in\mathbb{Z}\,,
\eea
and the asymptotic $\su(2)$ transfer matrix in the left and right sectors have the form \cite{Arutyunov:2010gu}
\bea
\la{asymptoticT}
T^{\su(2)\,(\ell,r)}_{Q,1}&=&(Q+1)\prod_{i=1}^{M}\frac{x^--x_i^-}{x^+-x_i^-}\sqrt{\frac{x^+}{x^-}}
-Q\, e^{-i \alpha_{\ell,r}}
\prod_{i=1}^{M}\frac{x^--x_i^+}{x^+-x_i^-}\sqrt{\frac{x^+x_i^-}{x^-x_i^+}}\\
 \nonumber
 &&-Q\, e^{i\alpha_{\ell,r}}
\prod_{i=1}^{M}\frac{x^--x_i^-}{x^+-x_i^-}\frac{x_i^--\frac{1}{x^+}}{x_i^+-\frac{1}{x^+}}
\sqrt{\frac{x^+x_i^+}{x^-x_i^-}}+(Q-1)\prod_{i=1}^{M}\frac{x^--x_i^+}{x^+-x_i^-}\frac{x_i^--\frac{1}{x^+}}{x_i^+-\frac{1}{x^+}}
\sqrt{\frac{x^+}{x^-}}\, ,
\eea
where $M$ is the number of magnons and $x^\pm,\ x^\pm_i$ are the usual parameterizations of mirror and string rapidities.

We will restrict to the choice
\bea
\a_3=0,\quad\quad\a=\a_\ell=\a_r=-\frac{\a_2}{2}\,,
\eea
and it is immediate to obtain the Bethe-Yang equation
\bea
-1=Y_{1*}^\circ(u_k)\,,
\eea
from the analytic continuation of the asymptotic $Y_Q^\circ$ functions
\bea
\la{asymptYQ}
Y_Q^\circ(v)=e^{-J\tilde{\cal E}_Q(v)}\,T_{Q,1}^{(\ell)}\big(v,\{u_k\}\big)\,T_{Q,1}^{(r)}\big(v,\{u_k\}\big)\,\prod_{j=1}^MS^{Q1*}_{\sl(2)}(v,u_j)\,.
\eea
One then finds
\bea
-1= e^{ip_k
J}e^{i\alpha_2}\,\prod_{j=1}^M S_{\sl(2)}(u_k,
u_j)\left(\frac{x_k^--x_j^+}{x_k^+-x_j^-}\sqrt{\frac{x_k^+x_j^-}{x_k^-x_j^+}}\right)^2\,.
\eea
which can be rewritten using the explicit form of the S-matrix  and the total momentum quantization condition (\ref{twistedtotalmom}) as
\bea
\la{allloopalpha}
1= e^{ip_k(J+M)}\,e^{i\a_2}\,e^{-i\a_1}\prod_{j\neq k}^M \frac{u_k-u_j-2i}{u_k-u_j+2i}\sigma^{-2}(u_k,u_j)\,.
\eea

Applying this discussion to the family of the states of interest, for which $M=3$, $J=J_1=L-3$, $J_2=3$ and $J_3=0$, one finds 
\bea
\la{twistcondition}
3\g_1+(L-3)\g_2=0\,,
\eea
whereas the Bethe-Yang equations can be written simply as
\bea
1= e^{ip_k\,L}\,e^{2\pi i\,L\,\g_3}\prod_{j\neq k}^M \frac{u_k-u_j-2i}{u_k-u_j+2i}\sigma^{-2}(u_k,u_j)\,,
\eea
so that we can think of twist as being related to a deformation by
\bea
\phi=-2\pi\, L\,\g_3=\frac{1}{2}\frac{L}{L-3}\,\a\,.
\eea

It is also interesting to notice that, in the case $L=6$, the constraint (\ref{twistcondition}) is compatible with the choice
\bea
\g_1=\g_2=\g_3\,,
\eea 
which is the Leigh-Strassler deformation preserving $\mathcal{N}=1$ supersymmetry and dual to the Lunin-Maldacena background  \cite{Lunin:2005jy}. Furthermore, inspecting (\ref{asymptoticT}) one finds that, on a solution of (\ref{allloopalpha}), the explicit dependence on the deformation parameter drops from the asymptotic transfer matrix. As a result, many of the analytic properties of the asymptotic Y-functions will be essentially the same as in the untwisted case.

\subsection{Twisted Y-functions and their analytic properties}
\label{app:Yfunctions}

\begin{table}
\begin{center}
\begin{tabular}{|c|c|c|}
\hline Y${}^o$-function  & Zeroes   & Poles\\
\hline $Y_{M|w}$   & $r_{M\pm1}^{\lambda_{M\pm1}}$ &  \\
\hline $1+Y_{M|w}$ & $r_M^{{\lambda_{M}}\,-}\,,\ r_M^{{\lambda_{M}}\,+}$ & $u_2-(M+1)i/g\,,\ u_3+(M+1)i/g$  \\
\hline $Y_{1|vw}$ &$u_1\,,\ r_0^{\lambda_0}$ & \\
\hline $1+Y_{M|vw}$ && $u_2+(M+1)i/g\,,\ u_3-(M+1)i/g$ \\
\hline $Y_-$      & $u_2^-\,,\ u_3^+$ & $u_2^+\,,\ u_3^-$ \\
\hline $Y_+$       & & $u_1^-$\\
\hline $1-Y_-$   & $r_0^{\lambda_0\,-}\,,\ r_0^{\lambda_0\,+}$ & \\
\hline $1-Y_+$   &&\\
\hline $Y_1$   & $r_0^{\lambda_0}$&$u_2^{++}\,,\ u_3^{--}$ \\
\hline $Y_2$   & &$u_2^{+}\,,\ u_3^{-}$ \\
\hline $Y_Q\,, Q\ge 3$   & &$u_2+{i\ov g}(Q-1)\,,\ u_3-{i\ov g}(Q-1)$ \\
\hline
\end{tabular}
\end{center}
\la{tab:twistedroots}
\caption{Relevant roots and poles of asymptotic Y-functions for general $L$. The index $\lambda_M=1,...,\lambda_M^{\rm max}$ labels different roots, and $\lambda_M^{\rm max}$ depends on $L$.
}
\end{table}

The asymptotic transfer matrices in the antisymmetric representation (\ref{asymptoticT}), together with Bazhanov-Reshetikhin formula \cite{BR}, yield all of the $T_{QQ'}$.\footnote{For practical purposes it can be convenient to directly find $T_{1Q'}$ by a duality transformation as detailed in \cite{Arutyunov:2011uz} rather than from  Bazhanov-Reshetikhin formula.} From those, one finds the auxiliary Y-functions \cite{GKV09}
\bea
Y_{M|w}=\frac{T_{1,M}T_{1,M+2}}{T_{2,M+1}}\,,\quad
Y_-=-\frac{T_{2,1}}{T_{1,2}}\,,\quad
Y_+=-\frac{T_{2,3}T_{2,1}}{T_{1,2}T_{3,2}}\,,\quad
Y_{M|vw}=\frac{T_{M,1}T_{M+2,1}}{T_{M+1,2}}\,,~~~~~
\eea
whereas the asymptotic $Y_Q$ functions are given by (\ref{asymptYQ}). All are real analytic functions of the mirror rapidity. The relevant analytic properties of the full Y-functions can be found from inspecting their asymptotic counterparts at small $g$. Recall that in doing so, we will always consider the regime $\phi\lesssim g$.

In table \ref{tab:twistedroots} the meromorphic structure of Y-functions is schematized. A few remarks on how this scenario depends on $L$ are in order:
\begin{enumerate}
\item Auxiliary functions $Y_{M|w}$ and $Y_-$ satisfy quantization conditions at the shifted values of the (real) roots $\{r_M^{\lambda_M}\}_{\lambda_M=1,...,\lambda_M^{\rm max}}$, which by contour deformation trick will appear in the TBA equations. Their number $\lambda_M^{\rm max}$ and their position will depend on the value of $L$ under consideration.
\item As discussed, the form of the TBA equation and of the energy and momentum formulae will depend on whether the complex rapidities $u_{2,3}$ lie inside or outside the physical strip, which depends on $L$.
\item As seen in the previous appendix, the case $L=6$ is special in that it can be linked to a deformation which preserves more supersymmetry. As a result, the large-$u$ asymptotic of  $Y_Q(u)$ will be different depending on whether $L=6$ or not, which is consistent with the fact that the relation between the TBA length $L_{TBA}$ and $J$ is modified when all supersymmetry is broken \cite{Arutyunov:2010gu}.
\item It is worth pointing out that $Y_2$ has poles at $u_2^-,\,u_3^+$, which lie very close to the real line. As can be seen from (\ref{onelooptwistedsol}), in the limit $g\ll\phi\ll1$ their distance from the real line is of order $\phi^L$.
\end{enumerate}

\subsection{TBA equations for the twisted theory}
\label{app:TBA}
The TBA equations for the family of states of interest can be engineered by contour deformation trick, taking into account the analytic properties for the state at hand. We write them in a rather general form, by introducing terms ${\sf  D}_{*}$ that indicate the driving terms of a given equation that depend on the roots $\{r_M^{\lambda_M}\}_{\lambda_M=1,...,\lambda_M^{\rm max}}$, coming from $Y_{M|w}=-1$ or $Y_-=1$.

For concreteness, we consider a more involved case in which the complex rapidities lie (just) outside the analyticity strip (which is the case of $L=6$), and express TBA equation in terms of simplified and hybrid equations only. When the rapidities are inside the analyticity strip there is no need to consider the quantization of the roots of $1+Y_2$ and therefore $u_{2,3}^{(2)}$ drop out from all equations. We refer the reader to \cite{Arutyunov:2011mk} for a detailed discussion of the TBA equations with complex rapidities, whereas the definition of the kernels used below can be found in \cite{{AFS09}}.

\subsubsection*{Simplified equations for  $Y_{M|w}$}

\bea
\log Y_{M|w} &=&  \log(1 +  Y_{M-1|w})(1 +
Y_{M+1|w})\star s+ \delta_{M1}\, \log{1-{1\ov Y_-}\ov 1-{1\ov Y_+} }\hstar s +{\sf  D}_{M|w} .~~~~~
\eea
\subsubsection*{Simplified equations for  $Y_{M|vw}$}

\begin{align}
\log Y_{M|vw} = & - \log(1 +  Y_{M+1})\star s +
\log(1 +  Y_{M-1|vw} )(1 +  Y_{M+1|vw})\star
s\\
&  + \delta_{M1} \log{1-Y_-\ov 1-Y_+}\hstar s + \delta_{M1}\Big(
\log\frac{S(u_2^{(2)+} - v)}{S(u_3^{(2)-} - v)}- \log S(u_1^- - v)\Big)+{\sf  D}_{M|vw}\,\nonumber.
\end{align}

\subsubsection*{Simplified equations for  $Y_\pm$}
\bea
\log {Y_+\ov Y_-} &=& \,   \log(1 +  Y_{Q})\star K_{Qy}\\
&& - \sum_i \log S_{1_*y}(u_i^{(1)},v)+\log {S_{2y}(u_2^{(1)+}, v)\ov S_{2y}(u_2^{(2)+}, v)}{S_{2y}(u_3^{(2)-}, v)\ov S_{2y}(u_3^{(1)-}, v)}\,,\nonumber
\eea

\bea
\nonumber
 \log {Y_+ Y_-} &=&\  2\log {1+Y_{1|vw}
 \ov 1+Y_{1|w}}\star s- \log\left(1+Y_Q
\right)\star K_Q
+2\log(1 + Y_{Q}) \star K_{xv}^{Q1}\star s
 \\
 \nonumber
 &&
-\log {{S}_2(u_2^{(1)+}- v)\ov {S}_2(u_2^{(2)+}-v)}{{S}_2(u_3^{(2)-}- v)\ov {S}_2(u_3^{(1)-}- v)}
+2 \log {S^{21}_{xv}(u_2^{(1)+}, v)S^{21}_{xv}(u_3^{(2)-}, v)\ov S^{21}_{xv}(u_2^{(2)+}, v)S^{21}_{xv}(u_3^{(1)-}, v)} \star s
  \\
  &&-2 \log S_{xv}^{1_*1}(u_1,v)\star s+\log {S_2(u_1- v)} \star s
 \\
&& -2 \log {S_{xv}^{11}(u_2^{(1)},v)\ov S_{xv}^{11}(u_3^{(1)},v)}\star s+\log {{S}_1(u_2^{(1)}- v)\ov {S}_{1}(u_3^{(1)}- v)}+{\sf  D}_{+\times-}\,.
\nonumber
\eea
\subsubsection*{Simplified TBA equations for $Y_Q$}

\bigskip
 \noindent
$\bullet$ $\ Q\ge 4\ $
\bea
\log Y_{Q}&=&\log{\left(1 +  {1\ov Y_{Q-1|vw}} \right)^2\ov (1 +  {1\ov Y_{Q-1} })(1 +  {1\ov Y_{Q+1} }) }\star s
\,~~~~~~~
\eea

\bigskip
 \noindent
$\bullet$ $\ Q=3\ $

\bea
\log Y_{3}&=&\log S(u_2^{(2)+}-v)-\log S(u_3^{(2)-}-v)+\log{\left(1 +  {1\ov Y_{2|vw}} \right)^2\ov (1 +  {1\ov Y_{2} })(1 +  {1\ov Y_{4} }) }\star s
\,.~~~~~~~
\eea

\bigskip
 \noindent
$\bullet$ $\ Q=2\ $
\bea
\log Y_{2}&=&\log S(u_2^{(1)}-v)-\log S(u_3^{(1)}-v)+\log{\left(1 +  {1\ov Y_{1|vw}} \right)^2\ov (1 +  {1\ov Y_{1} })(1 +  {1\ov Y_{3} }) }\star_{p.v} s
\,,~~~~~~~
\eea

\subsubsection*{Hybrid TBA equations for $Y_Q$}
Following \cite{Arutyunov:2011mk} we introduce a function which combines the terms on the right hand side of the hybrid ground state TBA equation
\bea
G_Q(v)&=& - L_{\rm TBA}\, \tH_{Q} +\log \left(1+Y_{Q'} \right)
\star (K_{\sl(2)}^{Q'Q}+2 s\star K_{vwx}^{Q'-1,Q})\\
&+&  2 \log \(1 + Y_{1|vw}\) \star s \hstar K_{yQ} +2\log(1+Y_{Q-1|vw})\star s\nonumber \\
&  -&  2  \log{1-Y_-\ov 1-Y_+} \hstar s \star K^{1Q}_{vwx} +  \log
{1- \frac{1}{Y_-} \ov 1-\frac{1}{Y_+} } \hstar K_{Q}  +  \log
\big(1-\frac{1}{Y_-}\big)\big( 1 - \frac{1}{Y_+} \big) \hstar
K_{yQ} \, . \nonumber  \eea
Then the hybrid TBA equations for $Y_Q$ read 
\begin{align}
\log Y_Q(v) &= G_Q(v) -\log \frac{S_{\sl(2)}^{1Q}(u_2^{(1)},v)}{S_{\sl(2)}^{1Q}(u_3^{(1)},v)}S_{\sl(2)}^{1_*Q}(u_1,v)\nonumber+\log \frac{S_{\sl(2)}^{2Q}(u_3^{(2)-},v)}{S_{\sl(2)}^{2Q}(u_3^{(1)-},v)}\frac{S_{\sl(2)}^{2Q}(u_2^{(1)+},v)}{S_{\sl(2)}^{2Q}(u_2^{(2)+},v)} \\ \nonumber
& -\log
S^{1Q}_{vwx}(u_1,v) + 2\log S(u_1^-,v)\star_{p.v.} K_{vwx}^{1Q}
 \\
&
-2\log {S(u_2^{(2)+},v)\ov S(u_3^{(2)-},v)}\star K_{vwx}^{1Q}+{\sf  D}_{Q} \, . \end{align}

The exact Bethe equations can be found by analytic continuation of e.g. the hybrid equations to the string region. In the next appendix, we will consider them for the case $L=6$. 

\subsubsection*{Driving terms in the $L=6$ case}
The case on which we focus for explicit calculations is $L=6$. There, one has that there is always exactly one root $r_M$ for any $M$, so that the driving terms take the explicit form
\bea
{\sf  D}_{M|w}&=&- \log S(r_{M-1}^- - v)S(r_{M+1}^- - v)\,\\
\nonumber
{\sf  D}_{M|vw}&=& -\delta_{M1} \log S(r_{0}^- - v)\,\\
\nonumber
{\sf  D}_{+\times-}&=&+\log S(r_{1}^- - v)\,\\
\nonumber
{\sf  D}_{Q=1}&=&+\log S^{1Q}_{vwx}(r_0,v) -\log S_Q(r_0^--v)S_{yQ}(r_0^-,v)+2\log S(r_0^-,v)\star_{p.v.} K_{vwx}^{1Q}\,.
\eea
Since the twist preserves one supersymmetry, we have \cite{AFS09}
\bea
L_{TBA}=J+2\,.
\eea

\subsubsection*{Driving terms in the $L=10$ case}
As another example, we consider a state with $L=10$ for which rapidities are outside the analyticity strip. One finds that the auxiliary functions $Y_{M|w}$ and $Y_-$ satisfy quantization conditions at three distinct (shifted) rapidities $r^{(1)}_M,r^{(2)}_M,r^{(3)}_M$ for any $M$. As a result, the driving terms are now 
\bea
{\sf  D}_{M|w}&=&- \sum_{i=1}^3\log S((r^{(i)}_{M-1})^- - v)S((r^{(i)}_{M+1})^- - v)\,,\\
\nonumber
{\sf  D}_{M|vw}&=& -\delta_{M1}\sum_{i=1}^3 \log S((r^{(i)}_{0})^- - v)\,,\\
\nonumber
{\sf  D}_{+\times-}&=&\sum_{i=1}^3\log S((r^{(i)}_{1})^- - v)\,,\\
\nonumber
{\sf  D}_{Q=1}&=&\sum_{i=1}^3\left[\log S^{1Q}_{vwx}\left(r^{(i)}_{0},v\right) -\log S_Q\left((r^{(i)}_{0})^--v\right)S_{yQ}\left((r^{(i)}_{0})^-,v\right)\right]\\
\nonumber
&& +2\sum_{i=1}^3\log S\left((r^{(i)}_{0})^-,v\right)\star_{p.v.} K_{vwx}^{1Q}\,.
\eea
Furthermore, in this case we have
\bea
L_{TBA}=J\,.
\eea

\subsection{Linearized TBA and exact Bethe equations for $L=6$}
\label{app:linearizedTBA}
To find the first perturbative correction to the asymptotic quantization conditions it is convenient to expand the TBA system and exact Bethe equations around their asymptotic solution. As discussed, this will leave us with  three equations (\ref{expandedEBE}), two of which are complex and conjugate to each other, in three real unknowns $\delta u_1^{(1)}$, ${\rm Re}(\delta u_2^{(1)})$ and ${\rm Im}(\delta u_2^{(1)})$. These equations are compatible with the quantization of total momentum (\ref{momen3}). This allows one to find a solution for $\delta u^{(1)}_i$ by considering one of the two complex exact Bethe equations together with (\ref{momen3}).

To this end, we consider the exact Bethe equation for $u_2^{(1)}$, that is
\begin{align}
&\log(-1)=\log Y_{1}(u_2^{(1)}) = G_1(u_2^{(1)})
+  2 \log \(1 + Y_{1|vw}\) \star \tilde{s}  -\log \frac{S_{\sl(2)}^{11}(u_2^{(1)},u_2^{(1)})}{S_{\sl(2)}^{11}(u_3^{(1)},u_2^{(1)})}
S_{\sl(2)}^{1_*1}(u_1,u_2^{(1)}) \nonumber \\
&-2\log {S(u_2^{(2)+}, u_2^{(1)})\ov S(u_3^{(2)-}, u_2^{(1)})}\star
K_{vwx}^{11} -\log S_1(r_0^--u_2^{(1)})S_{y1}(r_0^-,u_2^{(1)})
\nonumber \\
&+\log \frac{
S_{\sl(2)}^{21}(u_3^{(2)-},u_2^{(1)})
}
{     S_{\sl(2)}^{21}(u_3^{(1)-},u_2^{(1)})  }
+\log \frac{{\rm Res}\, S_{\sl(2)}^{21}(u_2^{(1)+},u_2^{(1)})}
{S_{\sl(2)}^{21}(u_2^{(2)+},u_2^{(1)}) \,{\rm Res}\, Y_2(u_2^{(1)+})} \label{ExactB2} \\
& +2\log{\rm Res}\, S\star K_{vwx}^{11}(u_1^-,u_2^{(1)})- \log\Big(u_1-u_2^{(1)}-\frac{2i}{g}\Big)^2
\left(\frac{x_s^-(u_1)-\frac{1}{x^-(u_2^{(1)})}}{x_s^-(u_1)-\frac{1}{x^+(u_2^{(1)})}}\right)^2\nonumber\\
& +2\log {\rm Res}\, S\star K_{vwx}^{11}(r_0^-,u_2^{(1)})-\log\Big(r_0-u_2^{(1)}+\frac{2i}{g}\Big)^2
\left(\frac{x_s^+(r_0)-x^+(u_2^{(1)})}{x_s^+(r_0)-x^-(u_2^{(1)})}\right)^2 \, . \nonumber\end{align}
where we used the fact that $u_2$ lies in the overlap of string and mirror regions, and introduced the short-hand notation
\bea
\log {S(u_2^{(2)+}, u_2^{(1)})\ov S(u_3^{(2)-}, u_2^{(1)})}\star
K_{vwx}^{11}\equiv\int dt\log {S(u_2^{(2)+},t)\ov S(u_3^{(2)-},t)}\,
K_{vwx}^{11}(t, u_2^{(1)})\,.
\eea

We now want to expand this and the other TBA equations, by considering
\bea
Y_{M|w}(v)&=&Y_{M|w}^\circ (v)(1+ \mathscr{Y}_{M|w}(v))\frac{S(r^{\circ-}_{M-1}-v)}{S(r^{-}_{M-1}-v)}\frac{S(r^{\circ-}_{M+1}-v)}{S(r^{-}_{M+1}-v)},\ \ \ M\geq 1\\\nonumber
Y_{1|vw}(v)&=&Y_{1|vw}^\circ (v)(1+ \mathscr{Y}_{1|vw}(v))\frac{S(r^{\circ-}_{0}-v)}{S(r^{-}_{0}-v)},\\\nonumber
Y_{M|vw}(v)&=&Y_{M|vw}^\circ (v)(1+ \mathscr{Y}_{M|vw}(v)),\ \ \ M\geq 2\\
Y_{\pm}(v)&=&Y_{\pm}^\circ (v)(1+ \mathscr{Y}_{\pm}(v))\frac{S(r^{-}_{1}-v)}{S(r^{\circ-}_{1}-v)}\,.\nonumber
\eea  
Here $Y^\circ_*(v)$ are computed out of the asymptotic transfer matrices evaluated at the exceptional rapidities $u_i^{(1)}$. These vanish at some root $r_*^\circ=r_*^\circ(u_i^{(i)})$ that is not the exact root $r_*$ dictated by the quantization conditions coming from TBA. The S-matrices on the right hand side have  poles at these roots, so that the corrections $\mathscr{Y}_*$ are always small on the real line. For any Y-function it is convenient to introduce 
\bea
A_*=\frac{Y^\circ_*}{1+Y^\circ_*}\,.
\eea
Since in many equations  terms involving $u_{2,3}^{(2)}$ occur, we will have to consider their variation. In particular, it is convenient to express them in terms of the difference between $u_{2,3}^{(1)}$ and $u_{2,3}^{(2)}$, which we will indicate as
\bea
\delta u_{2,3}\equiv u_{2,3}^{(2)}-u_{2,3}^{(1)}\,,
\eea
and for which we know an asymptotic expression (\ref{deltauexpansion}). This quantity should not be confused with the corrections $\delta u_{2,3}^{(1)}$ which are the quantities that we are looking for,  and which cannot be found from asymptotic considerations. In a similar way, we also write
\bea
\delta r_M\equiv r_M-r^\circ_M\,.
\eea

We now proceed expanding the TBA equations.

\subsubsection*{Expansion of $Y_{M|w}$ equations}
\bea
\nonumber
\mathscr{Y}_{M|w}&=&A_{M-1|w}\left(\mathscr{Y}_{M-1|w}-2\pi i\,s(r^\circ_{M-2}-v)\delta r_{M-2}-2\pi i\,s(r^\circ_{M}-v)\delta r_{M}\right)\star s\\
\nonumber
&+&A_{M+1|w}\left(\mathscr{Y}_{M+1|w}-2\pi i\,s(r^\circ_{M}-v)\delta r_{M}-2\pi i\,s(r^\circ_{M+2}-v)\delta r_{M+2}\right)\star s\\
&+&\delta_{M1}\left(-\frac{A_-}{Y_-^\circ}\mathscr{Y}_-+\frac{A_+}{Y_+^\circ}\mathscr{Y}_+-2\pi i\left(\frac{A_-}{Y_-^\circ}\mathscr{Y}_--\frac{A_+}{Y_+^\circ}\right)s(r_1^\circ-v)\delta r_1\right)\hat{\star}s\,.
\eea

\subsubsection*{Expansion of $Y_{M|vw}$  equations}
\bigskip
 \noindent
$\bullet$ $\ M=1\ $
\bea
\nonumber
\mathscr{Y}_{1|vw}&=&A_{2|vw}\mathscr{Y}_{2|vw}\star s+\left(-A_-\mathscr{Y}_{-}+A_+\mathscr{Y}_{+}-2\pi i s(r^\circ_1-v)(A_--A_+)\,\delta r_1\right)\hat{\star} s\\
&+&2\pi i\,s(u_2^+-v)\delta u_2-2\pi i\,s(u_3^--v)\delta u_3-Y^\circ_2\star s\,.
\label{linearYVW1}
\eea

\bigskip
 \noindent
$\bullet$ $\ M=2\ $
\bea
\nonumber
\mathscr{Y}_{2|vw}&=&A_{1|vw}\mathscr{Y}_{1|vw}\star s+A_{3|vw}\mathscr{Y}_{3|vw}\star s-2\pi i\,A_{1|vw} s(r^\circ_0-v)\,\delta r_0\star s\\
&-&Y^\circ_3\star s\,.
\label{linearYVW2}
\eea

\bigskip
 \noindent
$\bullet$ $\ M\geq3\ $
\bea\label{linearYVWm}
\mathscr{Y}_{M|vw}&=&A_{M-1|vw}\mathscr{Y}_{M-1|vw}\star s+A_{M+1|vw}\mathscr{Y}_{M+1|vw}\star s-Y^\circ_{M+1}\star s\,.
\eea

\subsubsection*{Expansion of $Y_{\pm}$  equations}
\bigskip
 \noindent
$\bullet$ $\ \log Y_+/Y_-\ $
\bea
\label{linearYpm}
\mathscr{Y}_{+}-\mathscr{Y}_{-}&=&Y_Q^\circ\star K_{Qy}-2\pi i\,K_{2y}(u_2^{\circ,+},v)\delta u_2+2\pi i\,K_{2y}(u_3^{\circ,-},v)\delta u_3
\eea

\bigskip
 \noindent
$\bullet$ $\ \log Y_+Y_-\ $
\bea\nonumber
\mathscr{Y}_{+}+\mathscr{Y}_{-}&=&2A_{1|vw}\left(\mathscr{Y}_{1|vw}-2\pi i\,s(r^\circ_0-v)\delta r_0\right)\star s\\
\nonumber
&-&2A_{1|w}\left(\mathscr{Y}_{1|w}-2\pi i\,s(r^\circ_0-v)\delta r_0-2\pi i\,s(r^\circ_2-v)\delta r_2\right)\star s\\
\nonumber
&-&Y^\circ_Q\star K_Q+2Y^\circ_Q\star K^{Q1}_{vx}\star s+2\pi i\, K_2(u_2^+-v)\delta u_2-2\pi i\, K_2(u_3^--v)\delta u_3\\
&-&4\pi i \left(K^{21}_{xv}(u_2^+,v)\delta u_2-K^{21}_{xv}(u_3^-,v)\delta u_3\right)\star s.
\eea

\subsubsection*{Expansion of the quantization condition for $r_0$}
Since $r_0$ appears explicitly in the exact Bethe equation (\ref{ExactB2}), it will be necessary to consider its quantization condition. The quantization condition for $r_0$ should be found by continuing the equation for $Y_-$ to $-i/g$. One can however check that (\ref{linearYpm}) is subleading in $g$, so that we can directly work with the equation for $\log Y_+ Y_-$ and continue this down to $-i/g$. We have
\bea
 \log {Y_+^- Y_-^-} &=&\  2\log {1+Y_{1|vw}
 \ov 1+Y_{1|w}}\star_{pv} \tilde{s}- \log\left(1+Y_Q\right)\star_{pv} K^-_Q
 \\
 \nonumber
&&+\log {S_2(u_1- v)}\star_{pv} \tilde{s}+\log {{S}_1(u_2^{(1)+}- v)\ov {S}_{1}(u_3^{(1)+}- v)}
-\log {{S}_2(u_2^{(1)++}- v)\ov {S}_2(u_2^{(2)++}-v)}{{S}_2(u_3^{(2)}- v)\ov {S}_2(u_3^{(1)}- v)}
  \\\nonumber
 &&+\log S(r_{1} - v)+\log {1+Y_{1|vw}(v)
 \ov 1+Y_{1|w}(v)}-\frac{1}{2}\log\left(1+Y_1(v)\right)\,,
\eea
where we dropped all the contributions of kernels sub-leading in $g$. Evaluating this equation at $r_0$ yields a quantization condition, which can be expanded as follows:
\bea\nonumber
0&=&2A_{1|vw}\left(\mathscr{Y}_{1|vw}-2\pi i\,s(r^\circ_0-v)\delta r_0\right)\star \tilde {s}\\
\label{r0quantization}
&-&2A_{1|w}\left(\mathscr{Y}_{1|w}-2\pi i\,s(r^\circ_0-v)\delta r_0-2\pi i\,s(r^\circ_2-v)\delta r_2\right)\star \tilde {s}\\
\nonumber
&-&Y^\circ_Q\star_{pv} K_Q^-+2\pi i\, K_2(u_2^{++}-r_0^\circ)\delta u_2-2\pi i\, K_2(u_3-r_0^\circ)\delta u_3-\frac{1}{2}Y_1(r_0^\circ)\\
&+&\delta r_0\Big[2\log {1+Y^\circ_{1|vw}
 \ov 1+Y^\circ_{1|w}}\star_{pv} \tilde{s}'- \log Y^\circ_Q\star_{pv} (K^-_Q)'+\log {S_2(u_1- v)}\star_{pv} \tilde{s}'
\nonumber\\
& &+\log {S_2(u_1- v)}\star_{pv} \tilde{s}'+K_1(u_2^{+}- r_0^\circ)- K_{1}(u_3^{+}- r_0^\circ)+K(r_{1}^\circ - r_0^\circ)+\partial_vY^\circ_{1|vw}(r_0^\circ)
\nonumber\\
& &\ \ -
 \partial_vY^\circ_{1|w}(r_0^\circ)-\frac{1}{2}\frac{\partial_v Y^\circ_1(r_0^\circ)}{1+ Y_1(r_0^\circ)}+\frac{1}{2}{K_2(u_1- r_0^\circ)}\Big]\,,
\nonumber
\eea
where the primes denote derivatives with respect to the argument where $r_0$ is inserted.

\subsubsection*{Expansion of exact Bethe equation for $u_2$}
From the expansion of the exact Bethe equation we will be able to find the form of $\delta\mathcal{R}_{(2)}$, as outlined in (\ref{expandedEBE}). Some care is needed in dealing with the expansion of
\bea
\log \frac{{\rm Res}\, S_{\sl(2)}^{21}(u_2^{(1)+},u_2^{(1)})}
{S_{\sl(2)}^{21}(u_2^{(2)+},u_2^{(1)}) \,{\rm Res}\, Y_2(u_2^{(1)+})} =\log \frac{{\rm Res}\, S_{\sl(2)}^{21}(u_2^{(1)+},u_2^{(1)})}
{{\rm Res}S_{\sl(2)}^{21}(u_2^{(2)+},u_2^{(1)})}\frac{u_2^{(1)}-u_2^{(2)}}{{\rm Res}\, Y_2(u_2^{(1)+})} 
\eea
that according to (\ref{deltauexpansion}) can be written as
\bea
-2\pi i\,{\rm Res} K_{\sl(2)}^{21}(u_2^{\circ+},u_2^{\circ})\,\delta u_2-\frac{\partial {\rm Res Y_2}}{\partial u}\big(u_2^{\circ+}\big)\,.
\eea
The remaining terms can be readily expanded. Since we are interested in the lowest order correction to the quantization condition, we can also drop any sub-leading contribution in $g$, and in particular the terms containing the convolution $\hstar$. This leaves us with the final result
\bea
\nonumber
\delta\mathcal{R}_{(2)}&=&Y_Q^\circ\star\left(K^{Q1}_{\sl(2)}+2s\star K^{Q-1,1}_{vwx}\right)+2A_{1|vw}\left(\mathscr{Y}_{1|vw}-2\pi i\,s(r^{\circ-}_0-v)\delta r_0\right)\star\tilde s\\
\nonumber
&-&4\pi i \left(s(u_2^{\circ+},u_2^{\circ})\delta u_2-s(u_3^{\circ-},u_2^{\circ})\delta u_3\right)\star K_{vwx}^{11}\\
\nonumber
&+&2\pi i\,K_{\sl(2)}^{21}(u_3^{\circ-},u_2^{\circ})\,\delta u_3-2\pi i\,{\rm Res}{K}_{\sl(2)}^{21}(u_2^{\circ+},u_2^\circ)\,\delta u_2-\frac{\partial {\rm Res Y_2}}{\partial u}\big(u_2^{\circ+}\big)\\
&+&4\pi i\,{\rm Res}{s}\star K^{11}_{xvw}(r_0^\circ,u_2^\circ)\delta r_0-2\pi i\mathcal{K}(r_0^\circ,u_2^\circ)\delta r_0\,,
\la{deltar2}
\eea
where have introduced the notation $\mathcal{K}(u,v)=\frac{1}{2\pi i}\frac{d}{du}\mathcal{S}(u,v)$ with
\bea
\mathcal{S}(u,v)=S_1(u^--v)S_{y1}(u^-,v)\left(u^{--}-v\right)^2\left(\frac{x^+_s(u)-x^+_s(v)}{x^+_s(u)-x^-_s(v)}\right)^2\,,
\eea
in order to conveniently group all driving terms involving $r_0$.

\subsubsection*{Cancellation of the most $\phi$-divergent terms}
Finding an explicit expression for the contribution of $\delta\mathcal{R}_{(2)}$ to $\delta u^{(1)}_2$ and in turn to the seven-loop energy is highly non-trivial. The task is much more complicated than in the case of the Konishi operator \cite{AFS10,BH10a} because in (\ref{deltar2}) the correction $\delta r_0$ appears explicitly, together with $\mathscr{Y}_{1|vw}$. To determine these, one would have to solve both linear system associated to $\mathscr{Y}_{M|vw}$ and to $\mathscr{Y}_{M|vw}$, together with the equation yielding the quantization condition for $\delta r_0$. All these are coupled which makes finding a solution, even numerically, a complicated task.

For the purpose of finding evidence of a non-trivial cancellation of the divergent terms in the energy at $g^{14}$, however,  a much simpler analysis suffices.

Let us consider the $\cO(g^{12})$ part of (\ref{deltar2}) and of the linearized TBA equations, and expand them in powers of $\phi$. This expansion is expected to involve negative powers, which should be the ones that cure the divergences in the energy and that will come multiplying the sources of the linear systems.

For instance, in (\ref{linearYVW1}) the sources are
\bea
\nonumber
&&2\pi i\,s(u_2^+-v)\delta u_2-2\pi i\,s(u_3^--v)\delta u_3-Y^\circ_2\star s\\
&=&-2\pi i{\rm Res}Y_2^\circ(u_2^{\circ+})+2\pi i{\rm Res}Y_2^\circ(u_3^{\circ-})-Y^\circ_2\star s=\cO(\phi^{-6})\,,
\eea
due to the pole of $Y_2$ at $u_3^{\circ-}\approx u_2^{\circ+}+\cO(\phi^6)$. This implies that we can expect that $\mathscr{Y}_{1|vw}=\cO(\phi^{-6})$. Carrying out a  similar analysis for all the remaining TBA equations and quantization conditions for auxiliary roots, one  concludes that indeed $\mathscr{Y}_{1|vw}=\cO(\phi^{-6})$ and $\delta r_M=\cO(\phi^0)$.

Turning now to $\delta\mathcal{R}_2$, we find that up to higher orders in $\phi$ we have
\bea
\delta\mathcal{R}_{(2)}&=&Y_2^\circ\star K^{21}_{\sl(2)}(u_2^\circ)+2\pi i\,K_{\sl(2)}^{21}(u_3^{\circ-},u_2^{\circ})\,\delta u_3-\frac{\partial {\rm Res Y_2}}{\partial u}\big(u_2^{\circ+}\big)+\cO(\phi^{-6})\,.
\la{deltar2exp}
\eea
These three terms are all divergent at $\cO(\phi^{-12})$ due to the singularities of $Y_2$ and $K^{21}_{\sl(2)}$, and their contribution can be immediately evaluated in terms of asymptotic formulae. Inserting this into (\ref{expandedEBE}) and using that $P=\cO(\phi)$, one finds indeed that the most divergent part of the asymptotic energy at $g^{14}$, which goes like $\cO(\phi^{-8})$, is precisely canceled by wrapping corrections in (\ref{energyTBA}).


\end{document}